\documentclass[reprint,superscriptaddress,aps,prb,twocolumn]{revtex4-2}
\usepackage{setspace}
\usepackage{amsmath}
\usepackage{graphicx}
\usepackage{subfig}
\usepackage{ragged2e}
\usepackage{amsfonts}
\usepackage{amssymb}
\usepackage{epstopdf} 
\usepackage{xcolor}
\usepackage{verbatim}
\usepackage{soul}
\usepackage{comment}
\usepackage{placeins}
\usepackage{textcomp}
\usepackage{bm}
\usepackage{hyperref}
\usepackage{xcolor}
\usepackage[normalem]{ulem} 
\usepackage{makecell}
\usepackage{array}
\usepackage{booktabs}
\usepackage{caption}

\DeclareGraphicsExtensions{.pdf,.eps,.png,.jpg,.mps} 

\usepackage{float}
\renewcommand{\figurename}{\textbf{Figure}}

\begin{document}

\title{Overcoming noise-agility trade-off in integrated lasers for precision sensing}

\author{
Di Yu$^{1,*}$, Yitian Tong$^{1,*}$, Yu Xia$^{1}$, Yuntao Zhu$^{2}$, Yuemin Li$^{1}$, Mingfei Liu$^{1}$, Zhaoting Geng$^{1}$, Yuhao Huang$^{1}$, Yaoran Huang$^{1}$, Zheng Li$^{1}$, Jie Wang$^{1}$, Yunqi Fu$^{1}$, Hongjie Liang$^{1}$, Hao Fang$^{1}$, Jinwen Lin$^{1}$, Xuewen Chen$^{1}$, Kang Li$^{1}$, Xinlun Cai$^{3,\dagger}$ \& Chao Xiang$^{1,\dagger}$\\
$^1$Department of Electrical and Computer Engineering and State Key Laboratory of Optical Quantum Materials, The University of Hong Kong, Hong Kong, China\\
$^2$Liobate Technology Co., Ltd., Nanjing, Jiangsu 210003, China\\
$^3$State Key Laboratory of Optoelectronic Materials and Technologies, School of Electronics and Information Technology, Sun Yat-sen University, Guangzhou 510275, China\\
$^*$These authors contributed equally to this work. \\ 
$^\dagger$Corresponding authors: caixlun5@mail.sysu.edu.cn, cxiang@eee.hku.hk}

\begin{abstract}

Lasers that combine narrow linewidths with rapid tunability are critical for applications such as coherent optical ranging, distributed fiber-optic sensing, and precision spectroscopy. Despite significant progress in integrated laser technologies, the concurrent realization of low phase noise and frequency agility on a single integrated platform remains challenging owing to a fundamental architectural trade-off: conventional integrated laser designs typically suppress phase noise via high-$Q$ resonators, yet the extended photon lifetimes inherent to such resonators intrinsically constrain tuning speed. Here, we address this noise-agility trade-off by introducing a laser architecture that achieves ultralow phase noise and ultrafast tunability simultaneously. Rather than relying on ultrahigh-$Q$ resonators for self-injection locking, our design employs strong synthetic feedback within a Pockels-tunable, resonator-enhanced distributed Bragg reflector to suppress phase noise. As a proof of concept, we demonstrate a hybrid integrated laser with a short-term linewidth of 29 Hz, realized using a lithium niobate external cavity with a loaded $Q$ of only 0.62 million. The adoption of a moderate resonator $Q$ relaxes the photon-lifetime constraint on tuning speed, enabling sub-exahertz-per-second tuning rates and a chirp nonlinearity as low as 0.14\%. Leveraging this laser, we implement a frequency-modulated continuous-wave LiDAR system that achieves a relative ranging precision of $1.7 \times 10^{-4}$ at a measurement rate of 1 MSa\,s\textsuperscript{-1}, without requiring complex chirp linearization techniques. We further demonstrate fiber-optic acoustic sensing capable of detecting sub-$\mu\epsilon$ dynamic strain, underscoring the platform's versatility for high-speed precision optical measurements. By integrating ultralow noise and ultrafast tunability within a manufacturable platform, our work provides a route toward scalable, high-performance sensing systems.

\end{abstract}

\maketitle

\section{Introduction}

Lasers that simultaneously achieve low phase noise and frequency-agile tunability are essential for a wide range of optical sensing and metrology systems, including coherent optical ranging~\cite{snigirev2023ultrafast,behroozpour2016electronic,riemensberger2020massively}, distributed fiber-optic sensing~\cite{cheng2023chip,wissmeyer2018looking}, and high-resolution spectroscopy~\cite{corato2023widely}. In these applications, low laser phase noise improves measurement precision and sensitivity~\cite{bianconi2025requirements}, while a high frequency-sweep rate enables faster data acquisition and enhanced dynamic response. Traditionally, achieving both attributes has required bulky laser systems that rely on complex external modulation and noise-compensation schemes~\cite{li2022analysis,satyan2009precise}. As compact and cost-effective alternatives, integrated lasers offering both low phase noise and high frequency agility have attracted considerable interest. Significant progress has recently been made in this direction. Pockels-tunable self-injection-locked (SIL) lasers have demonstrated sub-kHz linewidths together with petahertz-per-second tuning rates~\cite{li2023high}, while piezoelectrically tunable SIL lasers have achieved hertz-level linewidths~\cite{voloshin2025monolithic}. These approaches exploit the Pockels effect in lithium niobate (LN) and the stress-optic effect in silicon nitride (SiN), respectively, to enable agile laser-frequency control. More recently, electro-optically tunable extended distributed Bragg reflector (E-DBR) lasers have further extended the achievable tuning speed to the exahertz-per-second regime~\cite{siddharth2025ultrafast,xue2025pockels}.

\begin{figure*}[t!]
    \centering
    \includegraphics{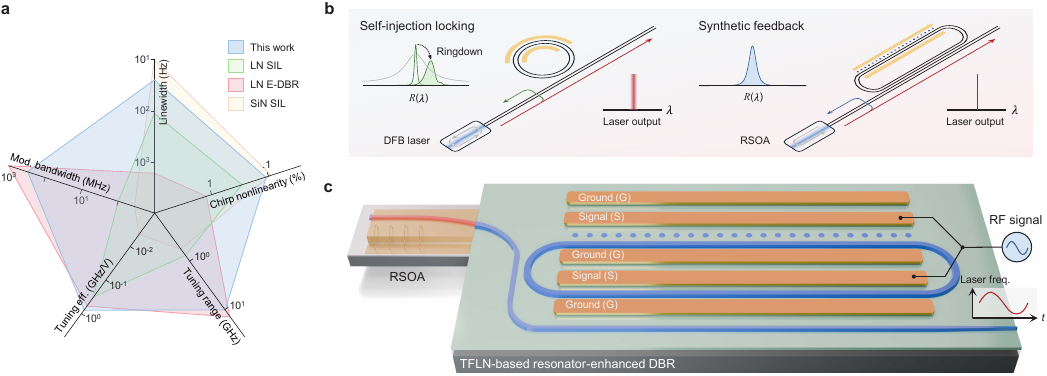}
    \captionsetup{singlelinecheck=off, justification = RaggedRight}
    \caption{
        \textbf{Laser performance trade-offs and the RE-DBR laser architecture.}
        \textbf{a}, Performance trade-space of representative integrated laser platforms. The performance polygon illustrates the set of metrics that can be simultaneously achieved in a laser architecture, highlighting the trade-offs among different laser properties, including short-term linewidth, modulation bandwidth, chirp nonlinearity, tuning efficiency, and mode-hop-free tunability.
        \textbf{b}, Conceptual diagrams illustrating high-speed modulation in a self-injection-locked laser employing a high-$Q$ resonator (left) and in a RE-DBR laser with a moderate-$Q$ external cavity (right). In the high-$Q$ case, resonator ringdown leads to spectral broadening and resonance shift of the feedback signal, which degrades laser coherence under rapid modulation.
        \textbf{c}, Schematic of the RE-DBR laser. An RSOA is butt-coupled to a TFLN-based grating-assisted resonator, forming a hybrid integrated laser that combines narrow linewidth with frequency-agile tuning. The electrode configuration is compatible with a GSGSG RF probe and enables high-speed electro-optic frequency tuning.
    }
    \label{fig:laser_architecture}
\end{figure*}

Despite these advances, it remains difficult to achieve ultralow phase noise and ultrafast tunability simultaneously in integrated lasers due to architecture-dependent trade-offs. This limitation constrains their use in demanding sensing systems such as precision LiDAR~\cite{roos2009ultrabroadband} and sensitive fiber-optic vibration detection~\cite{arbel2014dynamic,li2020high}. To clarify these trade-offs, Fig.~\ref{fig:laser_architecture}a maps the performance trade-space of several representative integrated laser designs, emphasizing noise characteristics and wavelength tunability~\cite{li2023high,siddharth2025ultrafast,xue2025pockels,voloshin2025monolithic}. While E-DBR lasers offer frequency agility and broad tunability, their phase noise performance is compromised, with typical short-term linewidths at the kHz level. Further linewidth reduction generally requires increasing the cavity photon lifetime by using longer external-cavity Bragg gratings~\cite{xiang2019ultra,yu2026resonator}, an approach constrained by fabrication non-uniformities in these sub-wavelength structures and, in some cases, by material anisotropy~\cite{li2022integrated,siddharth2025ultrafast}. In contrast, self-injection-locked lasers can achieve ultra-narrow fundamental linewidths down to several hertz by employing ultrahigh-$Q$ resonators~\cite{jin2021hertz,puckett2021422,xiang20233d,zhang2017monolithic,zhu2024twenty,nishimoto2026photonic}, but their modulation bandwidth is limited by the long photon lifetime intrinsic to such resonators. Additionally, the attainable frequency tuning rate is restricted by resonator ringdown dynamics, which degrade laser stability and coherence when the tuning rate surpasses the squared resonance linewidth, leading to a trade-off between low phase noise and fast frequency agility~\cite{siddharth2025ultrafast} (left panel of Fig.~\ref{fig:laser_architecture}b; see also Supplementary Note C).

In this work, we address the noise-agility trade-off by introducing an integrated laser architecture that simultaneously achieves ultralow phase noise and ultrafast frequency tunability. The architecture uses a resonator-enhanced distributed Bragg reflector (RE-DBR) as an external cavity, providing Pockels-tunable synthetic feedback that can be engineered for both frequency agility and narrow linewidth. Theoretically, we show that the RE-DBR external-cavity laser can achieve a linewidth comparable to that of self-injection-locked lasers while requiring a substantially lower resonator $Q$. This significantly relaxes the high-speed modulation constraints that typically limit conventional narrow-linewidth integrated lasers. Experimentally, we implement a hybrid integrated RE-DBR laser on a thin-film lithium niobate platform and achieve a short-term linewidth of 29 Hz---a record for integrated Pockels lasers---with a loaded resonator quality factor of only 0.62 million. Enabled by this moderate quality factor, the laser achieves an electro-optic tuning rate of up to $0.25$~EHz\,s\textsuperscript{-1}, a modulation nonlinearity as low as 0.14\%, and a mode-hop-free tuning range exceeding 10 GHz. These properties make the device a compelling source for high-precision coherent optical sensing. Leveraging the RE-DBR laser, we demonstrate both a frequency-modulated continuous-wave (FMCW) LiDAR system and a fiber-optic acoustic sensing platform. The FMCW LiDAR achieves a relative ranging precision of $1.7\times10^{-4}$ at a measurement rate of 1 MSa\,s\textsuperscript{-1}, while the fiber-optic sensing system resolves sub-$\mu\epsilon$ strain in optical fibers. Notably, both demonstrations operate without phase-noise compensation or chirp linearization, substantially simplifying system complexity. Together, these results establish the RE-DBR laser as a promising integrated platform for advanced optical sensing and metrology applications.

\section{Laser architecture}

The Pockels-tunable RE-DBR laser combines a reflective semiconductor optical amplifier (RSOA) with a grating-assisted racetrack resonator, which is built on thin-film lithium niobate (TFLN) and serves as an external cavity (see Fig.~\ref{fig:laser_architecture}c). When the grating's Bragg wavelength matches a racetrack resonance, the external cavity produces strong, narrowband reflection, effectively acting as a single-wavelength mirror. Under this condition, the laser achieves single-mode, narrow-linewidth operation~\cite{yu2026resonator,reep2025compact}. To exploit the Pockels effect in lithium niobate for electro-optic tuning, electrodes are positioned next to the resonator. These electrodes are designed as lumped capacitors rather than high-bandwidth transmission lines, which avoids modulation-induced frequency mismatch between counter-propagating resonant modes and enhances the stability of single-mode operation.

Notably, the RE-DBR laser overcomes the architecture-dependent performance trade-offs that limit conventional narrow-linewidth integrated lasers, such as E-DBR and self-injection-locked lasers. In an E-DBR laser, narrow-linewidth operation demands a long, uniform Bragg grating to deliver the required narrowband feedback~\cite{xiang2019ultra}. Achieving such large-scale uniformity is challenging on integrated platforms due to fabrication imperfections. The RE-DBR laser circumvents this issue through a resonator enhancement effect that extends the grating's effective length beyond its physical dimensions. This extension produces narrowband reflection and enables low-phase-noise operation within a compact footprint. SIL lasers, on the other hand, achieve injection locking through random Rayleigh scattering in a high-$Q$ resonator; a high quality factor is essential for both stable operation and linewidth reduction~\cite{kondratiev2017self}, which in turn limits their frequency agility. In contrast, the RE-DBR laser uses an external cavity with engineerable feedback strength and reflection bandwidth. By employing strong synthetic feedback, it can reach ultralow phase noise without requiring a high-$Q$ resonator, thereby overcoming the trade-off between narrow linewidth and frequency agility inherent to SIL lasers (see Supplementary Note A).

\begin{figure*}[t!]
    \centering
    \includegraphics{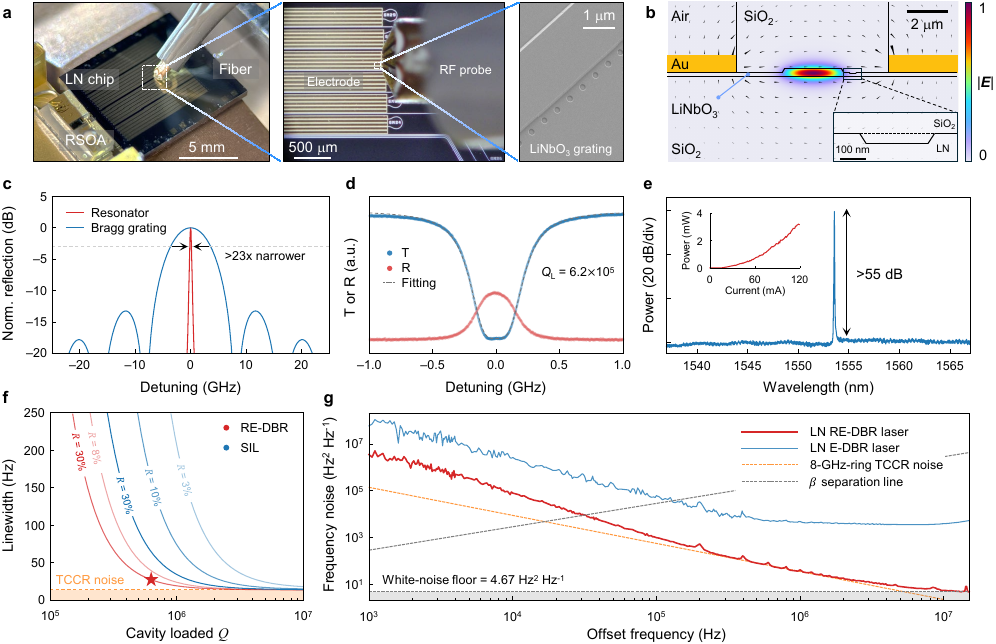}
    \captionsetup{singlelinecheck=off, justification = RaggedRight}
    \caption{
        \textbf{Experimental demonstration of the Pockels-tunable RE-DBR laser.}
        \textbf{a}, Photograph (left), optical microscope image (middle), and SEM image (right) of the fabricated device.
        \textbf{b}, Simulated optical (color) and RF electric field (arrows) distribution across the resonator cross-section.
        \textbf{c}, Calculated reflection spectra for a standalone 8-mm-long Bragg grating and for a RE-DBR incorporating a grating with identical specifications.
        \textbf{d}, Measured transmission and reflection spectra of the RE-DBR near the peak reflectivity wavelength.  
        \textbf{e}, Optical spectrum of the RE-DBR laser. Inset: on-chip optical power as a function of injection current.
        \textbf{f}, Simulated short-term linewidth of the RE-DBR laser compared to a self-injection-locked (SIL) laser, plotted against the external-cavity loaded quality factor $Q$ and the feedback strength $R$. The red star marks the configuration used in this experimental demonstration.
        \textbf{g}, Single-sideband frequency noise power spectral density for the RE-DBR laser (red), an E-DBR laser (blue), and the TCCR noise in a TFLN ring resonator with an 8-GHz free spectral range (FSR). The RE-DBR laser reaches a white noise floor of 4.67 Hz\textsuperscript{2}\,Hz\textsuperscript{-1}, corresponding to a short-term linewidth of 29.3 Hz.
    }
    \label{fig:experimental_demo}
\end{figure*}

We experimentally demonstrate the Pockels-tunable RE-DBR laser by butt-coupling an indium phosphide (InP) RSOA to a TFLN-based RE-DBR external cavity. Figure~\ref{fig:experimental_demo}a shows a photograph, an optical microscope image, and a scanning electron microscope (SEM) image of the device. The RSOA is a commercially available component with a broadband gain spectrum centered at 1550 nm~\cite{PHOTONX}. The RE-DBR external cavity consists of a racetrack resonator integrated with a weakly coupled (low-$\kappa$) waveguide Bragg grating, which is formed by a periodic array of shallow-etched holes placed alongside the resonator. Gold electrodes are positioned above the resonator for probe contact and electro-optic tuning.

The TFLN external cavity chip is fabricated on a lithium-niobate-on-insulator wafer using a wafer-scale process (see Methods). The wafer consists of a 360-nm-thick X-cut lithium niobate layer on a silicon substrate, separated by a 9-$\mu$m-thick buried oxide. Fabrication of the RE-DBR structures involves multiple etching steps. First, ridge waveguides and resonators are defined by etching 180 nm into the lithium niobate layer. A second, shallower etch of 40 nm is then applied to the remaining 180-nm-thick slab to form the Bragg gratings. After further processing to create the edge couplers, the wafer is covered with an oxide cladding. In selected regions, this oxide is etched down to the lithium niobate layer to form grooves, which are subsequently filled with metal electrodes. The resulting resonator cross-section is shown in Fig.~\ref{fig:experimental_demo}b. In this figure, the optical axis of the lithium niobate lies parallel to the horizontal axis. The radiofrequency (RF) electric field is represented by arrows, while the electric field of the waveguide's fundamental mode is shown using a color gradient.

The racetrack resonator has a waveguide width of 2.8 $\mu$m and supports multiple transverse modes. Among these, the fundamental mode exhibits an intrinsic loss of approximately 9 dB\,m\textsuperscript{-1}, as extracted from the transmission spectrum of a racetrack resonator without Bragg gratings (Extended Data Fig.~\ref{fig:ring_transmission}). The resonator circumference is 19 mm, corresponding to a free spectral range (FSR) of 8 GHz for the fundamental mode. The Bragg grating used in the RE-DBR is 8 mm long and consists of a periodic array of holes etched into the lithium niobate slab, with a period of 851 nm. This geometry yields a second-order grating with a Bragg wavelength around 1550 nm. Although higher-order gratings are relatively easy to fabricate due to their large feature sizes, they inherently introduce radiation loss from grating diffraction~\cite{streifer1977coupled}. To suppress this radiation loss, the grating is implemented as a shallow-etched array of holes (300 nm diameter, 40 nm depth) rather than being fabricated together with the waveguides in a single etch step (Supplementary Note E). The grating is separated from the resonator by 375 nm, resulting in a grating coupling coefficient of $\kappa = 0.1$ cm\textsuperscript{-1}. Additionally, the coupling between the resonator and the bus waveguide is engineered to achieve a power coupling ratio of approximately 20\%. Numerical simulations indicate that this parameter set enables stable single-mode operation, a narrow laser linewidth, and moderate output power (see Supplementary Note D).

The resonator enhancement effect significantly increases the effective length of the grating, thereby sharpening the wavelength selectivity of its reflection. According to coupled mode theory calculations, this effect produces a synthetic reflection in our RE-DBR external cavity with a 3-dB bandwidth more than 23 times narrower than that of the grating alone, indicating the potential for ultralow laser phase noise (Fig.~\ref{fig:experimental_demo}c)~\cite{yu2026resonator}. To enable electro-optic tuning, each device includes five 900-nm-thick gold electrodes. These rectangular electrodes share identical dimensions and equal spacing, and are designed for compatibility with a 50 $\mu$m-pitch ground-signal-ground-signal-ground (GSGSG) RF probe, facilitating high-bandwidth connections to RF signal sources and measurement equipment. The electrode-to-waveguide separation is set to 2.25 $\mu$m to balance tuning efficiency against metal-induced absorption loss. To enable efficient edge-coupling to the RSOA and to a cleaved single-mode fiber, bilayer mode converters are integrated at both ends of the TFLN chip, following the design described in Ref.~\cite{wang2024high}. To reduce Fresnel reflection at the coupling facets, index-matching gel is applied to the fiber-to-chip coupling facet. After edge polishing, the TFLN chip achieves a fiber coupling loss as low as 1.2~dB per facet.

\section{Ultralow-phase-noise operation}

The transmission and reflection spectra of the RE-DBR external cavity are presented in Extended Data Fig.~\ref{fig:redbr_spectrum}. A dominant reflection peak centered at 1553.5 nm is observed, coinciding with a transmission dip that indicates a resonance. A magnified view of this spectral feature is shown in Fig.~\ref{fig:experimental_demo}d, revealing a peak reflectivity of 36\% and a reflection bandwidth of 340 MHz. Notably, this synthetic reflectivity provided by the RE-DBR exceeds the typical reflectivity arising from Rayleigh scattering in high-$Q$ resonators ($\sim$3\%) by an order of magnitude~\cite{xiang20233d,snigirev2023ultrafast}. To determine the quality factor of this resonance, the transmission spectrum was fitted with a doublet Lorentzian model~\cite{puckett2021422}, yielding a loaded $Q$ of 0.62 million. This value is relatively moderate and is readily achievable on thin-film lithium niobate (TFLN) platforms using wafer-scale fabrication processes~\cite{li2023high}.

We experimentally demonstrate single-mode, ultralow-phase-noise operation of the RE-DBR laser, leveraging these wavelength-selective reflection characteristics. Figure~\ref{fig:experimental_demo}e displays the optical spectrum of the laser, measured at an injection current of 120 mA, confirming single-mode emission with a side-mode suppression ratio (SMSR) exceeding 55 dB. The on-chip laser power as a function of injection current is shown in the inset of Fig.~\ref{fig:experimental_demo}e, revealing a threshold current of 16 mA and a maximum output power of 3.2 mW. Figure~\ref{fig:experimental_demo}g presents the measured single-sided frequency noise power spectral density of the RE-DBR laser, obtained using the correlated self-heterodyne method~\cite{yuan2022correlated} (see Methods). For comparison, the frequency noise spectrum of an E-DBR laser employing an 8-mm-long TFLN Bragg grating as its external cavity is also shown (see Supplementary Note I for characterization results of the grating). At offset frequencies below 10 kHz, the frequency noise of the RE-DBR laser is dominated by technical sources, attributed to coupling instability and driving electronics noise. Between 200 kHz and 2 MHz, the noise approaches the fundamental thermal-charge-carrier-refractive (TCCR) noise floor, simulated for a TFLN ring resonator with an 8-GHz free spectral range (FSR)~\cite{zhang2025fundamental}. At a 10-MHz offset, the frequency noise reaches a white-noise floor of 4.67 Hz\textsuperscript{2}\,Hz\textsuperscript{-1}, corresponding to a short-term linewidth of 29.3 Hz.

Theoretical analysis was conducted to elucidate the ultra-narrow-linewidth behavior observed in the RE-DBR laser, a feature that is nontrivial given the laser's relatively moderate resonator $Q$. Figure~\ref{fig:experimental_demo}f shows the theoretically calculated short-term linewidth as a function of external-cavity loaded $Q$ and feedback strength $R$. The model accounts for two primary noise sources: spontaneous emission and TCCR noise; full details of the model and parameter configuration are provided in Methods. For comparison, the calculated linewidth of a self-injection-locked laser is also included. The experimental configuration, marked by a star, aligns closely with the theoretical results at $R=30\%$. Furthermore, the calculations reveal that, owing to differences in feedback strength and overall architecture, achieving a comparable linewidth with a SIL laser ($R=3\%$) would require a sixfold increase in external-cavity $Q$. These results quantitatively establish that the RE-DBR laser operates in a regime that combines narrow linewidth and relatively low resonator $Q$, which is inaccessible to SIL lasers.

\section{Frequency-agile laser tuning}

\begin{figure*}[t!]
    \centering
    \includegraphics{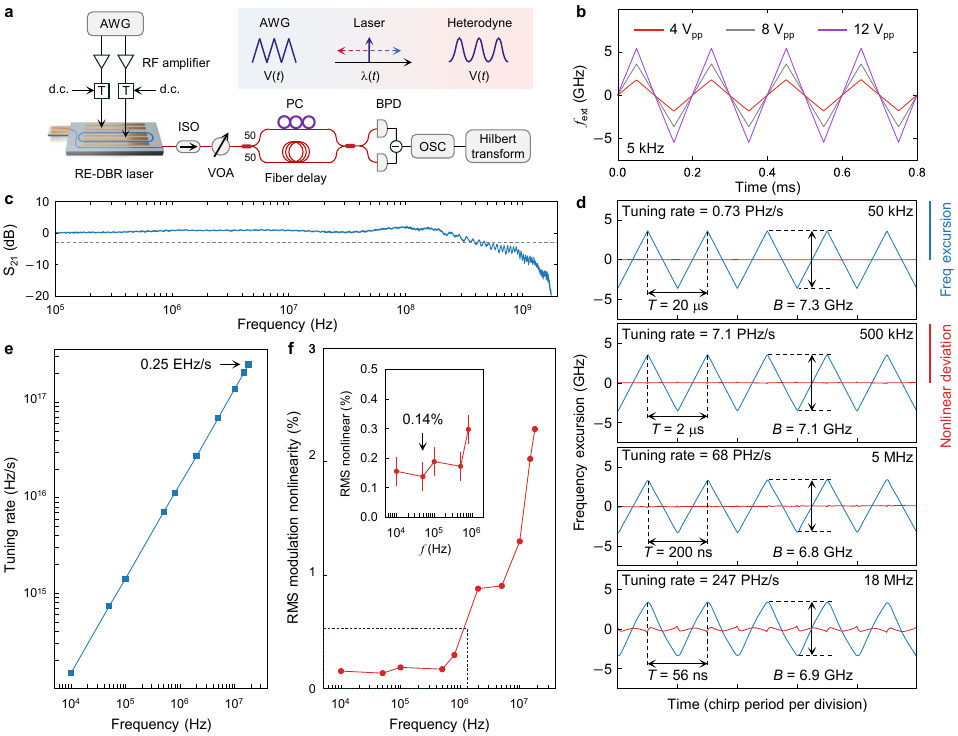}
    \captionsetup{singlelinecheck=off, justification = RaggedRight}
    \caption{
        \textbf{Electro-optic frequency tuning of the RE-DBR laser.}
        \textbf{a}, Schematic of the experimental setup for electro-optic tunability characterization. Two identical RF signals are applied to tune the laser, with the resulting frequency chirp characterized via delayed self-heterodyne detection. AWG, arbitrary waveform generator; ISO, isolator; VOA, variable optical attenuator; PC, polarization controller; BPD, balanced photodetector; OSC, oscilloscope.
        \textbf{b}, Measured laser frequency excursion under 5-kHz triangular-wave modulation.
        \textbf{c}, Small-signal modulation response of the RE-DBR external cavity, characterized as a microring intensity modulator with a vector network analyzer.
        \textbf{d}, Laser frequency spectrograms at various modulation frequencies, using two 8-V\textsubscript{pp} triangular-wave signals. Nonlinear frequency deviation is obtained by comparing the measured frequency excursion with an ideal triangular-wave fit. The corresponding frequency tuning rate (\textbf{e}) and RMS modulation nonlinearity (\textbf{f}) are plotted against modulation frequency. Inset: data for modulation frequencies below 1 MHz. Error bars indicate nonlinear frequency deviation from the measurement equipment.
    }
    \label{fig:frequency_agility}
\end{figure*}

We characterize the tunability of the RE-DBR laser using the experimental setup shown in Fig.~\ref{fig:frequency_agility}a. Electro-optic tuning is implemented by applying two identical RF signals to the GSGSG electrodes, resulting in a chirped laser output while maintaining the Bragg frequency aligned with the racetrack resonance. This alignment suppresses variations in the synthetic feedback strength and improves operational stability. We characterize the frequency-modulated laser output using delayed self-heterodyne detection and record the resulting electrical beat note. The laser frequency excursion is extracted from the phase of this beat note via a Hilbert transform, enabling analysis of tuning efficiency, chirp linearity, and modulation bandwidth (see Methods). To directly assess the intrinsic linearity of the frequency modulation, no predistortion or active feedback is used during the characterization.

To measure the low-frequency modulation response of the RE-DBR laser, we applied two synchronized 5-kHz triangular-wave signals for wavelength tuning. The amplitudes of these signals were varied simultaneously during the measurement, and the resulting laser spectrograms are presented in Fig.~\ref{fig:frequency_agility}b. The frequency excursion increased linearly with the modulation signal amplitude. Specifically, for peak-to-peak tuning amplitudes of 4, 8, and 12 V\textsubscript{pp}, we observed continuous tuning ranges of 3.68, 7.29, and 10.93 GHz, respectively. These results demonstrate excellent modulation linearity, corresponding to a tuning efficiency of 0.91 GHz\,V\textsuperscript{-1}.

The high-speed modulation capability of the device is evaluated by measuring its small-signal response with a vector network analyzer (VNA). In this measurement, the RE-DBR external cavity functions as a microring intensity modulator, with probe light supplied by an independent tunable laser whose wavelength is positioned on the linear slope of a RE-DBR resonance (Supplementary Note J). As shown in Fig.~\ref{fig:frequency_agility}c, the measured modulation response remains flat up to 100 MHz and exhibits a 3-dB bandwidth of 389 MHz. This bandwidth is comparable to the spectral linewidth of the resonance (Fig.~\ref{fig:experimental_demo}d), indicating that it approaches the fundamental limit imposed by the photon lifetime~\cite{li2013ring}.

Leveraging its combination of high modulation bandwidth and wide mode-hop-free tunability, the Pockels-tunable RE-DBR laser is well-suited for generating high-speed, linear frequency chirps for coherent optical sensing. To demonstrate this capability, triangular-wave tuning signals with a fixed peak-to-peak amplitude of 8 V\textsubscript{pp} are applied at frequencies ranging from 10 kHz to 18 MHz. The resulting frequency spectrograms, along with the deviations from triangular-wave fits that quantify chirp linearity, are shown in Fig.~\ref{fig:frequency_agility}d. These results confirm that the laser can generate highly linear chirps with a 7-GHz frequency excursion at MHz-scale repetition rates. At modulation frequencies of 10 MHz and above, both tuning efficiency and linearity begin to degrade, which is attributed to the finite modulation bandwidth of the device (see Supplementary Note B). As shown in Fig.~\ref{fig:frequency_agility}e, the maximum achieved tuning rate is 0.25 exahertz per second. This value is 1.65 times the square of the resonance linewidth, indicating operation near the ringdown limit~\cite{siddharth2025ultrafast}. Furthermore, the RE-DBR laser exhibits excellent chirp linearity, achieving a root-mean-square (RMS) modulation nonlinearity below 0.3\% for modulation frequencies up to 800 kHz (Fig.~\ref{fig:frequency_agility}f). When the laser is modulated at 50 kHz, its RMS modulation nonlinearity reaches a minimum value of 0.14\%.

\section{High-performance coherent optical ranging}

\begin{figure*}[t!]
    \centering
    \includegraphics{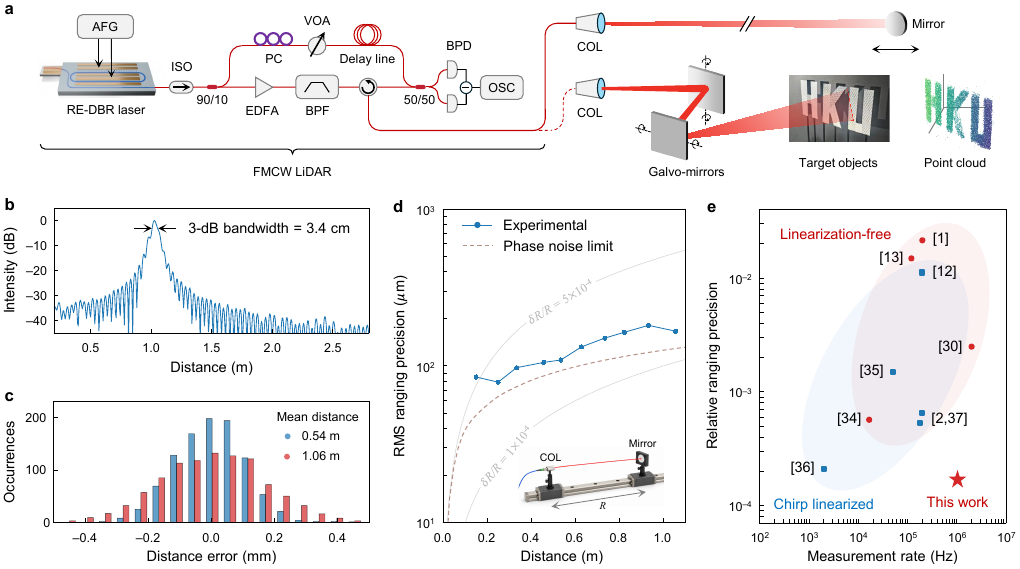}
    \captionsetup{singlelinecheck=off, justification = RaggedRight}
    \caption{
        \textbf{High-performance FMCW LiDAR.}
        \textbf{a}, Schematic of the experimental setup for FMCW LiDAR based on the RE-DBR laser. The system performance is evaluated by ranging a mirror mounted on an optical guide rail. Galvo-mirrors are used for beam scanning to enable three-dimensional imaging. AFG, arbitrary function generator; EDFA, erbium-doped fiber amplifier; BPF, band-pass filter; COL, collimator.
        \textbf{b}, Frequency spectrum of the LiDAR beat note. The spatial resolution is extracted from the 3-dB bandwidth of the signal.
        \textbf{c}, Statistical distribution of ranging results for a mirror placed at two different positions. The standard deviation of each distribution defines the RMS ranging precision.
        \textbf{d}, Ranging precision as a function of the mirror distance from the collimator. The combined phase noise of the laser and the RF source sets a fundamental lower bound on the achievable precision.
        \textbf{e}, Performance comparison of FMCW LiDAR systems employing integrated laser sources.
    }
    \label{fig:fmcw_lidar}
\end{figure*}

The combination of ultralow phase noise, ultrafast tunability, and excellent modulation linearity establishes the RE-DBR laser as a promising platform for coherent optical ranging. Using this laser, we demonstrate a FMCW LiDAR system that achieves both high ranging precision and a high measurement rate. The experimental setup is depicted in Fig.~\ref{fig:fmcw_lidar}a. The laser is frequency-modulated by a triangular waveform to generate a chirped, continuous-wave output. This chirped probe light is then employed in a delayed self-heterodyne detection scheme, producing an electrical beat note whose frequency encodes the target distance. The modulation frequency is set to 500 kHz, yielding a ranging measurement rate of 1 MSa\,s\textsuperscript{-1}, as both the upward and downward frequency ramps are used for distance measurement. The tuning signal has an amplitude of 8 V\textsubscript{pp}, corresponding to a total chirp bandwidth of 7 GHz. To mitigate ranging errors caused by modulation nonlinearity, only the central linear portion of each frequency ramp is utilized, reducing the effective frequency span to $B$ = 4.7 GHz and yielding a theoretical spatial resolution of $c/2B = 3.2$ cm (see Methods).

We benchmarked the coherent ranging system by measuring the distance to a mirror mounted on an optical guide rail. In the setup, a collimator couples the probe light from the optical fiber into free space. The mirror is positioned 1 m from the collimator, as indicated by the guide rail scale, and the resulting beat note is recorded. Its frequency spectrum, shown in Fig.~\ref{fig:fmcw_lidar}b, has been converted to a distance scale with zero-padding applied to enhance the spectral resolution. The spectral peak appears at approximately 1 m, in good agreement with the mirror-to-collimator distance read from the guide rail scale. Furthermore, the 3-dB bandwidth of the beat note corresponds to a spatial resolution of 3.4 cm for the LiDAR system, closely matching the theoretical limit.

The ranging precision of a LiDAR system is defined as the standard deviation of its distance measurements. Figure~\ref{fig:fmcw_lidar}c shows the statistical distributions of measurements at target distances of 0.56 m and 1.06 m, both of which demonstrate sub-millimeter precision. In general, the ranging precision of a FMCW LiDAR depends on target distance, as longer distances yield worse precision~\cite{behroozpour2016electronic}. To characterize this dependence in our system, we moved the mirror to various positions and performed approximately 1,000 measurements at each location (see statistics in Extended Data Fig.~\ref{fig:lidar_statistics}). From these data, we plot the ranging precision as a function of target distance, as shown in Fig.~\ref{fig:fmcw_lidar}d. As expected, the precision deteriorates at longer distances, a trend we attribute to low-frequency noise in the system. Notably, the measured precision approaches the theoretical lower bound determined by the combined phase noise of the RE-DBR laser and the RF source (see Supplementary Note G). This underscores the importance of a low-phase-noise source in a high-precision ranging system. At a target distance of 1 m, the system achieves a ranging precision of 170 $\mu$m, corresponding to a relative precision of $1.7\times10^{-4}$.

Figure~\ref{fig:fmcw_lidar}e compares the relative ranging precision and measurement rate of our FMCW LiDAR with those of other systems employing integrated laser sources~\cite{xue2025pockels,siddharth2025ultrafast,wang2024high,snigirev2023ultrafast,yang2023high,behroozpour2016electronic,lukashchuk2024photonic,liu2024highly,liu2025fast}. Among FMCW LiDAR demonstrations that do not rely on chirp linearization, our system achieves the highest relative ranging precision. While chirp linearization techniques can also deliver comparable precision, they introduce additional system complexity and impose loop-bandwidth constraints that typically reduce the measurement rate by orders of magnitude~\cite{liu2024highly,han2026high}. By contrast, our approach leverages a high-performance laser source that combines frequency agility with low phase noise, offering a low-cost and robust solution that simultaneously delivers high ranging precision and a high measurement rate.

\begin{figure*}[t!]
    \centering
    \includegraphics{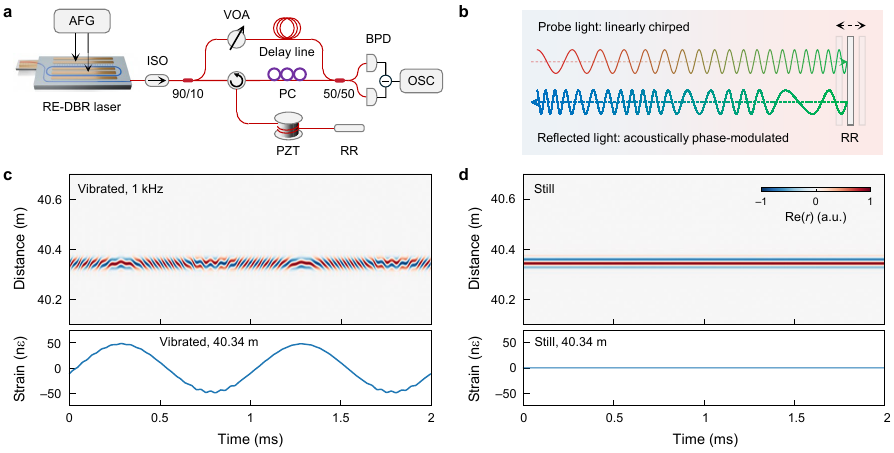}
    \captionsetup{singlelinecheck=off, justification = RaggedRight}
    \caption{
        \textbf{Fiber-optic acoustic sensing.}
        \textbf{a}, Schematic of the experimental setup for dynamic optical frequency domain reflectometry. A piezoelectric transducer (PZT) induces dynamic strain in a fiber under test terminated with a retroreflector (RR).
        \textbf{b}, Diagram of coherent optical sensing of acoustic vibrations. Linearly chirped probe light is launched into the fiber under test and reflected at the retroreflector. Dynamic strain modulates the round-trip optical path length, imprinting a phase shift on the reflected light.
        \textbf{c}, Measured complex reflectivity along the fiber (upper) and extracted fiber strain (lower) with 1-kHz PZT actuation (left).
        \textbf{d}, Corresponding results with the PZT on standby.
    }
    \label{fig:fiber_sensing}
\end{figure*}

\section{Fiber-optic acoustic sensing}

Fiber-optic sensing is another key application that demands frequency-agile, narrow-linewidth laser sources. The conventional approach relies on complex laser systems, such as fiber lasers combined with piezoelectric actuators and external single-sideband modulation~\cite{arbel2014dynamic,li2020high,zhu2025high}, resulting in high deployment costs that limit broader adoption. Here, we demonstrate that the Pockels-tunable RE-DBR laser can serve as an integrated, low-cost alternative for fiber-optic sensing. To validate this capability, we employ the laser in a fiber-optic acoustic sensing system based on optical frequency domain reflectometry (OFDR)~\cite{arbel2014dynamic}. The experimental setup is illustrated in Fig.~\ref{fig:fiber_sensing}a. The laser generates frequency-modulated probe light that travels through a 40-m-long fiber under test, where acoustic vibrations are introduced by a piezoelectric transducer (PZT). These vibrations induce dynamic strain along the fiber, which modulates the optical phase of the reflected light and gives rise to a time-varying complex reflectivity, as illustrated in Fig.~\ref{fig:fiber_sensing}b. The acoustically induced phase shift is extracted by demodulating the beat note obtained from delayed self-heterodyne detection (Supplementary Note H). This process reconstructs the complex reflectivity profile of the fiber under test, from which the dynamic strain is derived.

The complex reflectivity profile along the fiber under 1-kHz PZT actuation is shown in the upper panel of Fig.~\ref{fig:fiber_sensing}c. A pronounced reflectivity peak appears near 40.34 m, corresponding to the retroreflector mounted at the fiber end. The phase of the reflected light varies over time, indicating the presence of dynamic fiber strain. Phase demodulation at this peak location clearly reveals the 1-kHz vibration, as shown in the lower panel of Fig.~\ref{fig:fiber_sensing}c. We also extract the strain amplitude from the phase excursion, using a proportionality ratio determined by a calibration test. With a 1-V\textsubscript{pp} driving signal applied to the PZT, the measured peak-to-peak fiber strain amplitude is 100 n$\epsilon$, consistent with the PZT specification. This result verifies the sub-$\mu\epsilon$ sensitivity of our fiber-optic sensing system. For comparison, Fig.~\ref{fig:fiber_sensing}d presents the corresponding results without intentionally introduced strain, where the reflected phase remains constant and no dynamic strain is observed. Together with the preceding FMCW LiDAR experiment, this proof-of-concept demonstration of fiber-optic acoustic sensing establishes the RE-DBR laser as a versatile platform for coherent optical sensing systems.

\section{Discussion and outlook}

In this work, we have demonstrated a novel integrated laser architecture---the Pockels-tunable RE-DBR laser---that simultaneously achieves ultralow phase noise and ultrafast tunability. Combined with excellent chirp linearity and wide mode-hop-free tuning, these properties make the laser ideal for driving precision sensing systems. Moreover, the use of a higher-order Bragg grating and a moderate-quality-factor resonator ensures compatibility with wafer-scale production. As a result, the Pockels-tunable RE-DBR laser offers a cost-effective, compact alternative to the expensive, bulky high-performance lasers traditionally used in precision optical sensing and metrology, paving the way for low-cost implementations of precision sensing applications.

Looking ahead, our integrated laser can be leveraged and further enhanced to provide expanded functionality and improved performance. First, the Pockels-tunable RE-DBR laser could be combined with an electro-optic locking mechanism to suppress low-frequency phase noise, thereby improving its suitability for long-range coherent optical ranging and distributed fiber-optic sensing applications~\cite{wu2025electro}. Second, advanced integration techniques, such as heterogeneous integration~\cite{morin2024coprocessed,li2025heterogeneously} and micro-transfer printing~\cite{op2020heterogeneous,niels2026advances}, could be leveraged to improve the manufacturability and system stability of the RE-DBR laser, addressing the requirements for cost-effectiveness and robustness in potential commercial deployment. Third, the laser source could be integrated with additional optoelectronic components---such as optical amplifiers~\cite{cai2021erbium}, high-bandwidth photodetectors~\cite{desiatov2019silicon}, beam scanners~\cite{zhang2022large}, and driving electronics~\cite{lukashchuk2024photonic}---to enable fully integrated silicon photonics metrology systems, including chip-scale FMCW LiDAR engines and compact precision spectroscopy platforms that are potentially compatible with drone-based deployment. More broadly, integration with emerging TFLN-based electro-optic devices could unlock the full potential of frequency-agile, narrow-linewidth laser sources for precision sensing applications and beyond~\cite{hu2025integrated}.

\bibliography{main.bib}

\medskip

\vspace{3 mm}

\clearpage

\noindent\textbf{Methods}

\noindent \textbf{Device fabrication.} The TFLN external-cavity chip is fabricated on a 6-inch lithium-niobate-on-insulator wafer using a wafer-scale process. The wafer consists of a 360-nm-thick X-cut lithium niobate layer atop a 9-$\mu$m-thick buried oxide. The fabrication sequence proceeds in six steps. First, waveguide and resonator patterns are defined via deep ultraviolet (DUV) lithography and transferred into the lithium niobate with a 180-nm-deep dry etch. Second, a photolithography step is performed on the remaining 180-nm-thick lithium niobate, followed by a shallow 40-nm etch to create the periodic hole arrays that form the Bragg gratings. In the third photolithography step, tapered waveguides are patterned on the residual 180-nm-thick lithium niobate slab, forming the bilayer edge-coupler structures. Next, a 4.5-$\mu$m-thick oxide cladding is deposited by plasma-enhanced chemical vapor deposition. Then, electrode patterns are defined on the cladding oxide in a fourth photolithography step, after which the cladding is etched down to the lithium niobate. In the etched regions, 0.9-$\mu$m-thick gold electrodes are formed using a standard metal lift-off process. Finally, the wafer is diced and edge-polished to produce the external-cavity chips.

\medskip

\noindent \textbf{Laser linewidth measurement.} The frequency noise power spectral density of the laser was characterized using the correlated self-heterodyne technique described in Ref.~\cite{yuan2022correlated}. The experimental configuration consisted of an unbalanced fiber Mach-Zehnder interferometer (MZI) incorporating a 1-km fiber delay line, two balanced photodetectors (BPDs, Thorlabs PDB415C-AC), a 55-MHz acousto-optic modulator (AOM, Brimrose AMM-55-8-70-2FP), a polarization controller, and an oscilloscope (OSC, Siglent SDS7304A). The laser under test was directed through the MZI, and the output was split into two replicas using 50:50 fiber couplers, each subsequently measured by one of the two BPDs to produce two beat note signals. To extract the spectral density of the common-mode noise while rejecting uncorrelated photodetector noise and oscilloscope quantization noise, both signals were simultaneously recorded by the oscilloscope and their cross-correlation was computed. The frequency response of the MZI was de-embedded by applying a gain factor to the common-mode noise spectrum, yielding the laser's frequency noise spectrum. During these measurements, the laser was powered by a low-noise battery source (Newport LDX-3620B) to minimize extraneous technical noise. Additionally, the ambient temperature was stabilized at 25.00 $^{\circ}$C using a thermoelectric cooler controlled by a servo controller (Thorlabs TC300B).

\medskip

\noindent \textbf{Theoretical short-term laser linewidth.} The short-term linewidth of a TFLN-based RE-DBR laser comprises contributions from the spontaneous-emission-limited linewidth and the thermal-noise-limited linewidth. It can be expressed using the modified Schawlow-Townes formula~\cite{kazarinov2003relation,tran2019tutorial,yu2026resonator} together with the TCCR noise model~\cite{zhang2025fundamental} as

\begin{equation}
    \Delta\nu = \frac{\pi h\nu^{3}n_{sp}}{16PQ^{2}}(1+\alpha_{H}^{2})T\eta[\ln(\eta^{2}R)]^{2} + 2\pi S_{\nu,\rm{TCCR}}(f),
\end{equation}

where $h$ is Planck's constant, $\nu$ the lasing frequency, $n_{sp}$ the population inversion factor, $P$ the laser output power, $Q$ the loaded quality factor of the external cavity, $\alpha_{H}$ the amplitude-phase coupling factor, $R$ the external-cavity peak reflectivity, $T$ the external-cavity transmission at the wavelength of peak reflectivity, and $\eta$ the coupling efficiency between the gain element and the external cavity. $S_{\nu, \rm{TCCR}}(f)$ denotes the single-sideband frequency noise power spectral density of TCCR noise in the TFLN external cavity, evaluated at the offset frequency where the laser reaches the white frequency noise floor. In the theoretical analysis (Fig.~\ref{fig:experimental_demo}f), we assume $n_{sp}=2$, $P=1$ mW, $\nu=193.5$ THz, $\alpha_{H}=2.5$, $T=0.1$, and $\eta=-5$ dB. The TCCR noise level $S_{\nu,\rm{TCCR}}(f)$ is taken to be 2.2 Hz\textsuperscript{2}\,Hz\textsuperscript{-1} at a 10-MHz offset frequency for a TFLN-based racetrack resonator with an 8-GHz FSR (see Supplementary Note F).

For a self-injection-locked laser, the short-term linewidth is given by~\cite{ousaid2024low}

\begin{equation}
    \Delta\nu = \Delta\nu_{0}\frac{1+4\left(\frac{f}{\nu}\right)^{2}Q^{2}}{\left(1+\sqrt{R}\eta\frac{Q}{Q_{\rm LD}}\right)^{2}+4\left(\frac{f}{\nu}\right)^{2}Q^{2}} + 2\pi S_{\nu,\rm{TCCR}}(f),
\end{equation}

where $f$ is the offset frequency corresponding to the white noise floor in the frequency noise spectrum, $Q$ is the loaded quality factor of the external cavity, $Q_{\rm LD}$ is the loaded quality factor of the cold laser diode cavity, and $\Delta\nu_{0}$ is the free-running linewidth of the laser diode. Typical values for SIL lasers adopted in the numerical calculation are $\Delta\nu_{0}=100$ kHz, $f=10$ MHz, $\nu=193.5$ THz, $\eta=-5$ dB, and $Q_{\rm LD}=2.5\times10^{3}$~\cite{ousaid2024low}.

\medskip

\noindent \textbf{Characterization of high-speed modulation response.} The laser frequency excursion under high-speed modulation was extracted from the beat note signal obtained via delayed self-heterodyne detection~\cite{ahn2007analysis}. A Hilbert transformation was applied to the beat note signal to recover its instantaneous phase; since the laser frequency excursion is a linear function of this phase, it can be derived directly. To cover the wide range of modulation frequencies relevant to our characterization, two distinct measurement configurations were employed, each tailored to a different modulation frequency range. Both configurations used a high-bandwidth balanced photodetector (BPD, Thorlabs PDB480C-AC) and an oscilloscope (OSC, Keysight MSOX6004A). For modulation frequencies below 1 MHz, a two-channel arbitrary function generator (AFG, Keysight 33600A) provided the modulation signals, and a 10-meter fiber delay line was used in the heterodyne detection. For frequencies above 1 MHz, an arbitrary waveform generator (AWG, Keysight M8190A), together with external RF amplifiers and a DC bias, generated the modulation signals; in this case, the fiber delay line was shortened to 1 meter to accommodate the bandwidth limitations of the BPD and OSC.

\medskip

\noindent \textbf{FMCW LiDAR demonstration.} In the coherent optical ranging experiment, the RE-DBR laser was frequency modulated by electro-optic tuning using two identical triangular-wave signals generated by an arbitrary function generator (AFG, Keysight 33600A). Each signal had an amplitude of 8 V\textsubscript{pp} and a frequency of 500 kHz, producing a 7.1 GHz laser frequency sweep range. This frequency chirp generated a beat note signal whose frequency encoded the target position. Each frequency ramp lasted 1~$\mu$s, corresponding to a LiDAR measurement rate of 1 mega-sample per second (MSa\,s\textsuperscript{-1}). To avoid chirp nonlinearity near the turning points, only the middle two-thirds of each ramp was used for distance calculation. This time-windowing reduced the effective tuning range to $B = 4.7$~GHz, giving a theoretical spatial resolution of $\Delta x = c/2B = 3.2$~cm. The beat-note frequency, and thus the target distance, was extracted using the Hilbert transform. Specifically, the instantaneous phase of the beat note signal was first obtained as a function of time, and the time-averaged temporal derivative of this phase was then used to determine the signal frequency and corresponding target distance. Notably, when the received signal has a high signal-to-noise ratio (SNR), this method can achieve a ranging error of $\sigma \approx \Delta x/\sqrt{\rm SNR}$~\cite{buck2007high,behroozpour2017lidar}, which is much smaller than the transform-limited spatial resolution $\Delta x$, itself determined entirely by the signal duration~\cite{roos2009ultrabroadband}.

The in-fiber laser power delivered to the FMCW LiDAR was 1 mW. It was amplified to 20 mW using an erbium-doped fiber amplifier (EDFA, Amonics C+L band EDFA) and then reduced to 7.5 mW by a tunable bandpass filter (BPF, Alnair Labs BVF-100), which suppressed broadband spontaneous emission from the EDFA. After further losses in a fiber-optic circulator, the free-space output power after the collimator (COL, Thorlabs F220APC-1550) was 5.1 mW. Owing to its anti-reflection coating and in-collimator beam divergence, the collimator exhibited a high return loss of approximately 60 dB, which helped maintain a high signal-to-noise ratio in the measurements. A fiber delay line was also inserted into the fiber MZI. Its length was chosen such that the reflection from the collimator corresponded to a beat-note frequency close to zero. When a mirror served as the target, it was mounted on an optical guide rail and its orientation was adjusted at each position to maximize the reflected power collected by the collimator. Under these conditions, the mirror exhibited a return loss of 6 dB when the distance between the mirror and the collimator was less than 1 m. For both the mirror and the collimator, return loss was defined as the ratio between the in-fiber probe power and the reflected signal power.

We constructed a three-dimensional imaging system by integrating the optical ranging system with a two-dimensional beam-scanning module. Specifically, the beam scanner employed two galvo mirrors (Oeabt OGS-2AX25K) driven by triangular-wave signals at 100 Hz and 2 Hz for horizontal and vertical scanning, respectively, and the LiDAR measurement rate was set to 100 kSa\,s\textsuperscript{-1}. Using this setup, we captured a point cloud representation of an ``HKU'' letter target, as shown in Extended Data Fig.~\ref{fig:3d_imaging}. The reconstructed shape and orientation of the three letters agree well with the photograph in the inset, demonstrating the capability of our Pockels-tunable RE-DBR laser for FMCW LiDAR applications. The point cloud consists of 50,000 voxels and was acquired in 0.5 seconds. This acquisition rate is currently limited by the response speed of the galvo mirrors and could be substantially improved by employing micro-electromechanical systems (MEMS) for faster beam scanning~\cite{zhang2022large}.

\medskip

\noindent \textbf{Fiber-optic acoustic sensing experiment.} A fiber stretcher (Paulsson PZ2) is a device in which 40 m of polarization-maintaining fiber (PMF) is uniformly wound onto a piezoelectric transducer (PZT), and it is used to emulate an optical fiber subjected to mechanical perturbation. The PZT is driven by a 1 V\textsubscript{pp}, 1 kHz sinusoidal signal, which produces a fiber extension with a peak-to-peak amplitude of 3.8 $\mu$m, according to the device specification. To suppress the influence of polarization-dependent dispersion in the PMF, a Faraday mirror is placed at the end of the fiber stretcher. This arrangement ensures that the forward- and backward-propagating light fields have orthogonal polarization states, making the round-trip delay independent of the input polarization. The time-varying fiber strain induced by the PZT, representing acoustic vibration, is measured using linearly chirped probe light generated by the RE-DBR laser, following the method described in Ref.~\cite{arbel2014dynamic}. Specifically, the laser is modulated with a 500 kHz, 8 V\textsubscript{pp} triangular waveform, as in the FMCW LiDAR demonstration. The resulting chirped probe light is used in a delayed self-heterodyne detection scheme, in which the optical path displacement introduced by the fiber stretcher is converted into a phase shift of the beat-note signal. To extract the fiber strain, we apply a Fourier transform to the beat note and obtain the complex amplitudes of its spectral components, which correspond to the complex reflectivities at different positions along the fiber stretcher. Repeating this analysis for beat-note waveforms recorded during each laser frequency up-ramp produces a two-dimensional map of complex reflectivity versus fiber position and time. At any given position, the dynamic fiber strain and the time-varying phase of the complex reflectivity share a linear relationship---differing only by a constant scaling factor and a fixed offset (see Supplementary Note H). These two constants can be determined by applying a fiber strain of known amplitude and performing a calibration.

\medskip

\noindent \textbf{Performance comparison of narrow-linewidth tunable laser sources.} We present a quantitative benchmark of state-of-the-art tunable narrow-linewidth lasers, evaluating their noise performance and high-speed modulation capabilities. The results are summarized in Extended Data Table~\ref{tab:laser_performance}. The comparison focuses on several key figures of merit: short-term linewidth, modulation bandwidth, mode-hop-free tuning range, chirp nonlinearity, and tuning efficiency. For a fair assessment, the chirp nonlinearity is evaluated at a modulation frequency near 1 MHz without applying chirp linearization techniques such as predistortion and electro-optic locking. The benchmark reveals that existing laser sources achieving short-term linewidths below 100 Hz typically exhibit limited frequency agility, with modulation bandwidths under 1 MHz~\cite{yu2026resonator,voloshin2025monolithic,Koheras}. In contrast, to the best of our knowledge, the RE-DBR laser is the first platform to simultaneously achieve a sub-100-Hz linewidth and a frequency modulation bandwidth exceeding 100 MHz, representing more than a substantial improvement in modulation speed over previous ultralow-noise lasers.

\medskip

\vspace{3 mm}

\noindent \textbf{Data Availability}
The data used to produce the plots within this work will be released on \textit{Zenodo} upon publication of this manuscript.

\vspace{1 mm}

\noindent \textbf{Code Availability}
The code used to produce the plots within this work will be released on \textit{Zenodo} upon publication of this manuscript.

\vspace{1 mm}

\noindent \textbf{Acknowledgments}
We thank the funding support from the National Key R\&D Program of China (2024YFA1409300), the Research Grants Council of Hong Kong (C7143-25Y, N\_HKU774\_25, T46-705/23-R, STG3/E-704/23-N, STG3/E-104/25-N), the Innovation and Technology Commission of Hong Kong (GHP/230/22GD), the National Natural Science Foundation of China (6232290014), the Guangdong Provincial Quantum Science Strategic Initiative (GDZX2304004, GDZX2404002), and the Croucher Foundation.

\vspace{1 mm}

\noindent\textbf{Author Contributions}
The concept of this work was conceived by D.Y. and C.X. The structures were designed and the theoretical analysis was conducted by D.Y., with assistance from J.L., X.C. and K.L. The sample was fabricated by Y.Z. and X.C. Measurements were performed by D.Y. and Y.T., with assistance from Y.X., M.L., Z.G. and Y.H. The results were analyzed by D.Y. with assistance from Y.H., Z.L., J.W. and Y.F. D.Y. wrote the paper with input from Y.T. and Y.X. All authors commented on and edited the paper. The project was performed under the supervision of C.X.

\vspace{1 mm}

\noindent \textbf{Conflict of Interest} The authors declare no competing interests.

\vspace{1 mm}

\noindent \textbf{Author Information} Correspondence and requests for materials should be addressed to X.C. (caixlun5@mail.sysu.edu.cn) or C.X. (cxiang@eee.hku.hk).

\clearpage
\onecolumngrid
\renewcommand{\tablename}{\bf Extended Data Table}
\renewcommand{\figurename}{\bf Extended Data Fig.}
\setcounter{table}{0}
\setcounter{figure}{0}

\begin{figure}
    \centering
    \includegraphics{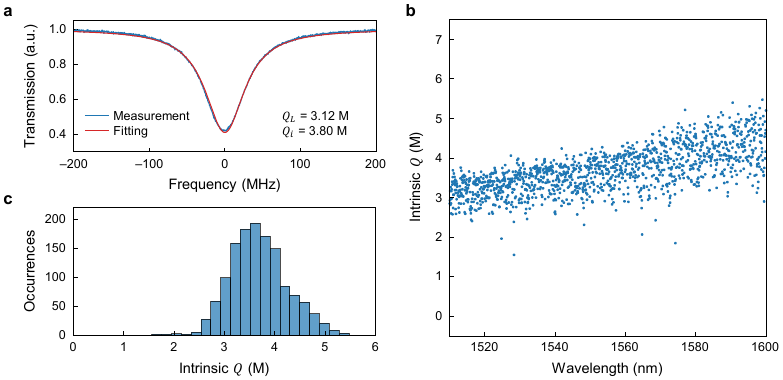}
    \captionsetup{singlelinecheck=off, justification = RaggedRight}
    \caption{
        \textbf{Characterization results of a racetrack resonator.} The resonator has a waveguide width of 2.8 $\mu$m and a circumference of 19 mm.
        \textbf{a}, Transmission spectrum of a resonance at 1550.313 nm. Fitting the dip with a Lorentzian line shape yields an intrinsic quality factor $Q_{i}$ of 3.8 million. Assuming a group index of 1.9 for the fundamental mode, this $Q_{i}$ corresponds to a waveguide loss of 9 dB\,m\textsuperscript{-1}.
        \textbf{b}, Extracted intrinsic quality factor as a function of wavelength. The intrinsic quality factor increases with wavelength, likely because optical modes at longer wavelengths exhibit weaker optical confinement, which reduces sidewall scattering and therefore waveguide loss.
        \textbf{c}, Statistical distribution of intrinsic quality factors. The $Q_{i}$ histogram reveals a single-peaked distribution indicative of single-transverse-mode operation. This observation contrasts with numerical simulations, which indicate that the 2.8-$\mu$m waveguide width is sufficient to support multiple transverse modes. We attribute this discrepancy to the substantially lower intrinsic quality factors of higher-order modes. In the present weakly coupled configuration, these low-$Q$ modes produce only shallow, nearly undetectable resonance dips, giving rise to the observed single-mode statistics.
    }
    \label{fig:ring_transmission}
\end{figure}

\clearpage

\begin{figure}
    \centering
    \includegraphics{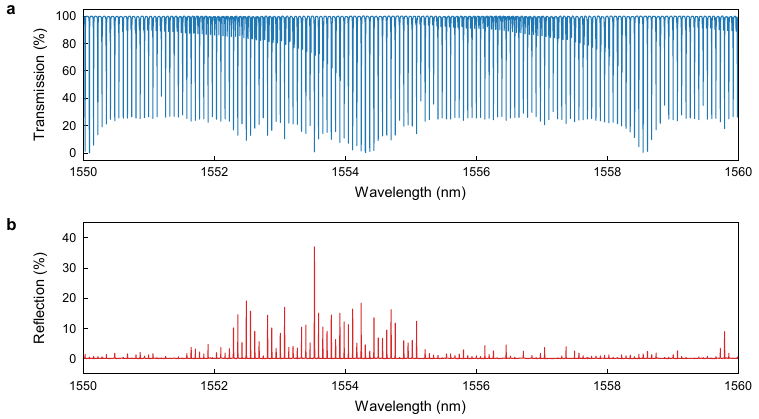}
    \captionsetup{singlelinecheck=off, justification = RaggedRight}
    \caption{
        \textbf{Transmission and reflection characteristics of the RE-DBR.}
        \textbf{a}, Measured transmission spectrum.
        \textbf{b}, Measured reflection spectrum, which exhibits a prominent peak centered at 1553.5~nm (a zoomed-in view of this peak is shown in Fig.~\ref{fig:experimental_demo}d). The minor reflection peaks appearing near the primary one are attributed to nonuniformities in the Bragg grating caused by fabrication imperfections.
    }
    \label{fig:redbr_spectrum}
\end{figure}

\clearpage

\begin{figure}[t!]
    \centering
    \includegraphics{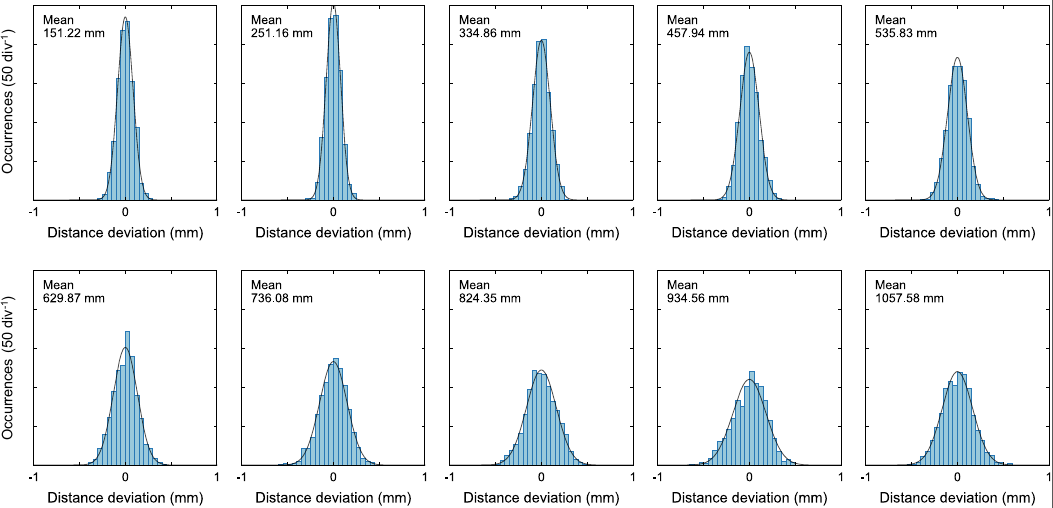}
    \captionsetup{singlelinecheck=off, justification = RaggedRight}
    \caption{
        \textbf{Histograms of FMCW LiDAR ranging results.} These panels show the statistical distribution of distance deviations when ranging a mirror placed at different positions. The black curve shows the Gaussian fit.
    }
    \label{fig:lidar_statistics}
\end{figure}

\clearpage

\begin{figure}[t!]
    \centering
    \includegraphics{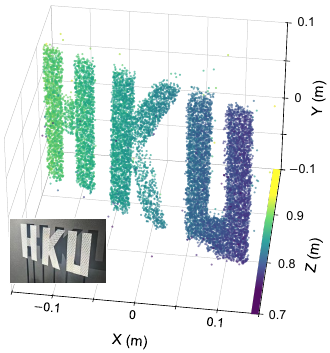}
    \captionsetup{singlelinecheck=off, justification = RaggedRight}
    \caption{
        \textbf{Three-dimensional point cloud representation.} Inset: the target object, consisting of three tilted letters. The z-coordinate denotes the distance from the fiber-optic collimator.
    }
    \label{fig:3d_imaging}
\end{figure}

\clearpage

\begin{table*}[htbp]
    \small
    \centering
    \begin{tabular}{cccccccc}
        \hline\hline
        
        \small\makecell{Laser\\Architecture} &
        \small\makecell{Short-term\\Linewidth (Hz)} &
        \small\makecell{External\\Cavity $Q$} &
        \small\makecell{Actuation\\Bandwidth (MHz)} &
        \small\makecell{Mode-hop-free\\Tunability (GHz)} &
        \small\makecell{RMS Chirp\\Nonlinearity} &
        \small\makecell{Tuning Effic.\\(MHz\,V\textsuperscript{-1})} &
        \small\makecell{Reference} \\
        \hline

        \multicolumn{8}{c}{\small\textbf{Integrated LiNbO\textsubscript{3} external-cavity lasers (electro-optic tuning)}} \\
        \hline
        RE-DBR & 29 & $6.2\times10^{5}$ & 389 & 10.9 & 0.17\% & 908 & \small\textbf{This Work} \\
        E-DBR & $2.8\times10^{3}$ & n.r. & 200 & 10 & 1.61\% & 550 & \cite{siddharth2025ultrafast} \\
        E-DBR & 167 & n.r. & $2\times10^{3}$ & 24 & 2.2\% & 800 & \cite{xue2025pockels} \\
        Vernier & $11.3\times10^{3}$ & $2.8\times10^{5}$ & 600 & 2 & 3\% & 340 & \cite{li2022integrated} \\
        Vernier & $9.42\times10^{3}$ & $2.1\times10^{4}$ & $>5$ & 3.44 & 0.11\% & 574 & \cite{wang2024high} \\
        Ring SIL & 163 & $1.2\times10^{6}$ & $<161^\dagger$ & 2.5 & 0.5\% & 380 & \cite{li2023high} \\
        \hline

        \multicolumn{8}{c}{\small\textbf{Integrated Si\textsubscript{3}N\textsubscript{4} external-cavity lasers (thermo-optic tuning)}} \\
        \hline
        RE-DBR & 24 & $5.6\times10^{5}$ & $<0.1$ & 34 & n.r. & n.r. & \cite{yu2026resonator} \\
        Ring SIL & $<9\times10^{3}$ & $5.5\times10^{4}$ & $<0.1$ & 33.9 & n.r. & n.r. & \cite{corato2023widely} \\
        E-DBR & $2.5\times10^{3}$ & n.r. & $<0.1$ & $>40$ & n.r. & n.r. & \cite{siddharth2024piezoelectrically} \\
        Vernier & $2.2\times10^{3}$ & n.r. & $<0.1$ & 28 & n.r. & n.r. & \cite{van2020ring} \\
        \hline

        \multicolumn{8}{c}{\small\textbf{Integrated Si\textsubscript{3}N\textsubscript{4} external-cavity lasers (stress-optic tuning)}} \\
        \hline
        Ring SIL & 3 & $5\times10^{6}$ & 0.4 & 0.5 & n.r. & 5 & \cite{voloshin2025monolithic} \\
        Ring SIL & 942$^\ddagger$ & $1.2\times10^{7}$ & 10 & 2 & 0.7\% & 520 & \cite{lihachev2022low} \\
        E-DBR & $2.5\times10^{3}$ & n.r. & 0.4 & 1 & 1.09\% & 4 & \cite{siddharth2024piezoelectrically} \\
        Vernier & 400 & $1.4\times10^{6}$ & 0.96 & 1.8 & $>1\%$ & 166 & \cite{lihachev2024frequency} \\
        \hline

        \multicolumn{8}{c}{\small\textbf{Integrated Si external-cavity lasers (thermo-optic tuning)}} \\
        \hline
        E-DBR & $1.1\times10^{3}$ & n.r. & $<0.1$ & n.r. & n.r. & n.r. & \cite{huang2019high} \\
        Vernier & $2.1\times10^{3}$ & n.r. & $<0.1$ & 31 & n.r. & n.r. & \cite{tran2019ring} \\
        Vernier & $5.7\times10^{3}$ & $2.5\times10^{4}$ & $<0.1$ & 375 & n.r. & n.r. & \cite{pintus2023demonstration} \\
        \hline

        \multicolumn{8}{c}
        {\small \textbf{Commercial low-noise erbium doped fiber laser (stress-optic tuning)}} \\
        \hline
        Fiber DFB & $<100$ & n.r. & $<0.02$ & 0.5 & n.r. & n.r. & \cite{Koheras} \\
        \hline\hline
        
    \end{tabular}
    \caption{Performance comparison of tunable narrow-linewidth lasers. $\dagger$Estimated from the spectral linewidth of the cavity resonance for self-injection locking. $^\ddagger$Data correspond to the device on which the frequency-agile tuning test was performed. The chirp nonlinearity is defined as the root-mean-square chirp frequency error divided by the chirp bandwidth, and it is evaluated at a modulation frequency of around 1 MHz. E-DBR: extended distributed Bragg reflector; SIL: self-injection locked; DFB: distributed feedback; n.r.: not reported.}
    \label{tab:laser_performance}
\end{table*}

\clearpage

\begin{table*}[htbp]
    \small
    \centering
    \setlength{\tabcolsep}{3pt}
    \begin{tabular}{cccccccc}
        \hline\hline
        \small\makecell{Laser Source} &
        \small\makecell{Actuation\\Mechanism} &
        \small\makecell{Ranging\\Precision} &
        \small\makecell{Target\\Distance} &
        \small\makecell{Relative\\Precision} &
        \small\makecell{Measurement\\Rate} &
        \small\makecell{Active Chirp\\Linearization} &
        \small\makecell{Reference} \\
        \hline
        LiNbO\textsubscript{3} RE-DBR & Electro-optic & 170 $\mu$m & 1 m & $1.7\times10^{-4}$ & 1 MSa\,s\textsuperscript{-1} & No & \small\textbf{This Work} \\
        LiNbO\textsubscript{3} E-DBR & Electro-optic & $>5.2$ mm$^\dagger$ & 0.35 m & $>1.5\times10^{-2}$ & 120 kSa\,s\textsuperscript{-1} & No & \cite{xue2025pockels} \\
        LiNbO\textsubscript{3} E-DBR & Electro-optic & 50 mm & 4.48 m & $1.1\times10^{-2}$ & 200 kSa\,s\textsuperscript{-1} & Predistortion & \cite{siddharth2025ultrafast} \\
        LiNbO\textsubscript{3} Vernier & Electro-optic & \makecell{4.9 mm\\10.9 mm} & \makecell{1.97 m\\3.64 m} & \makecell{$2.5\times10^{-3}$\\$3.0\times10^{-3}$} & 2 MSa\,s\textsuperscript{-1} & No & \cite{wang2024high} \\
        LiNbO\textsubscript{3}/Si\textsubscript{3}N\textsubscript{4} SIL & Electro-optic & 60 mm & 2.8 m & $2.1\times10^{-2}$ & 200 kSa\,s\textsuperscript{-1} & No & \cite{snigirev2023ultrafast} \\
        III-V DFB & Current injection & 390 $\mu$m & 0.689 m & $5.7\times10^{-4}$ & 16.7 kSa\,s\textsuperscript{-1} & No & \cite{yang2023high} \\
        III-V DBR & Current injection & \makecell{400 $\mu$m\\190 $\mu$m} & 0.36 m & \makecell{$1.1\times10^{-3}$\\$5.3\times10^{-4}$} & 180 kSa\,s\textsuperscript{-1} & \makecell{No\\Electro-optic PLL} & \cite{behroozpour2016electronic} \\
        Si\textsubscript{3}N\textsubscript{4} Vernier & Stress-optic & 15 mm & 10 m & $1.5\times10^{-3}$ & 50 kSa\,s\textsuperscript{-1} & Predistortion & \cite{lukashchuk2024photonic} \\
        Si\textsubscript{3}N\textsubscript{4} Vernier & Thermo-optic & 120 $\mu$m & 0.5 m & $2.4\times10^{-4}$ & 5 Sa\,s\textsuperscript{-1} & Predistortion & \cite{han2026high} \\
        Si\textsubscript{3}N\textsubscript{4} Vernier & Thermo-optic & \makecell{2.02 m\\42 mm} & 200 m & \makecell{$1.0\times10^{-2}$\\$2.1\times10^{-4}$} & 2 kSa\,s\textsuperscript{-1} & \makecell{Predistortion\\Electro-optic PLL} & \cite{liu2024highly} \\
        Si/Si\textsubscript{3}N\textsubscript{4} SIL & Plasma dispersion & 4 mm & 6.2 m & $6.5\times10^{-4}$ & 200 kSa\,s\textsuperscript{-1} & Predistortion & \cite{liu2025fast} \\
        \hline\hline
        
    \end{tabular}
    \caption{Performance comparison of FMCW LiDAR utilizing integrated laser sources. $^\dagger$Value estimated from spatial resolution. E-DBR: extended distributed Bragg reflector; SIL: self-injection locked; DFB: distributed feedback; PLL: phase-locked loop.}
    \label{tab:lidar_performance}
    
\end{table*}

\end{document}


\title{Supplementary Information for: Overcoming noise-agility trade-off in integrated lasers for precision sensing}

\author{
Di Yu$^{1,*}$, Yitian Tong$^{1,*}$, Yu Xia$^{1}$, Yuntao Zhu$^{2}$, Yuemin Li$^{1}$, Mingfei Liu$^{1}$, Zhaoting Geng$^{1}$, Yuhao Huang$^{1}$, Yaoran Huang$^{1}$, Zheng Li$^{1}$, Jie Wang$^{1}$, Yunqi Fu$^{1}$, Hongjie Liang$^{1}$, Hao Fang$^{1}$, Jinwen Lin$^{1}$, Xuewen Chen$^{1}$, Kang Li$^{1}$, Xinlun Cai$^{3,\dagger}$ \& Chao Xiang$^{1,\dagger}$\\
$^1$Department of Electrical and Computer Engineering and State Key Laboratory of Optical Quantum Materials, The University of Hong Kong, Hong Kong, China\\
$^2$Liobate Technology Co., Ltd., Nanjing, China\\
$^3$State Key Laboratory of Optoelectronic Materials and Technologies, School of Electronics and Information Technology, Sun Yat-sen University, Guangzhou 510275, China\\
$^*$These authors contributed equally to this work. \\ 
$^\dagger$Corresponding authors: caixlun5@mail.sysu.edu.cn, cxiang@eee.hku.hk}

\maketitle


\clearpage
\onecolumngrid
\renewcommand{\figurename}{\bf Extended Data Fig.}
\setcounter{figure}{0}

\tableofcontents

\newpage

\clearpage
\onecolumngrid
\appendix

\makeatletter
\@removefromreset{equation}{section} 
\makeatother
\renewcommand{\theequation}{S\arabic{equation}}
\renewcommand{\thefigure}{S\arabic{figure}}
\renewcommand{\thetable}{S\arabic{table}}
\setcounter{figure}{0}
\setcounter{equation}{0}
\setcounter{table}{0}

\section*{Supplementary Information}

\section{Trade-off between laser linewidth and frequency agility} \label{SI_tradeoff_linewidth_vs_tunability}

Narrow linewidth and fast frequency tuning are inherently conflicting laser properties. In this section, we quantitatively establish the trade-off between the intrinsic (spontaneous-emission-limited) linewidth and the maximum achievable tuning rate for three laser architectures: extended distributed Bragg reflector (E-DBR) lasers, self-injection-locked (SIL) lasers, and resonator-enhanced distributed Bragg reflector (RE-DBR) lasers. Our calculations indicate that, for a given linewidth, RE-DBR lasers can achieve higher frequency tuning rates than the other two conventional architectures.

The intrinsic linewidth of a semiconductor laser $\Delta\nu$ is given by the Schawlow-Townes formula as follows~\cite{kazarinov2003relation}

\begin{equation}  \label{eq:schawlow_townes_linewidth}
    \Delta\nu = \frac{\pi h\nu^{3}n_{sp}}{PQ_{l}^{2}}(1+\alpha_{H}^{2}),
\end{equation}

where $h$ is Planck's constant, $\nu$ the laser frequency, $n_{sp}$ the population inversion factor, $P$ the laser output power, $Q_{l}$ the loaded quality factor of the cold laser cavity, and $\alpha_{H}$ the amplitude-phase coupling factor.

Equation~\ref{eq:schawlow_townes_linewidth} assumes a two-mirror model in which light incident on a facet mirror is either reflected or transmitted. This assumption implies zero excess mirror loss, which does not hold for hybrid integrated lasers. In such lasers, an external cavity acts as one facet mirror and introduces non-negligible excess losses, including butt-coupling loss and on-chip loss. The other facet mirror, by contrast, has high reflectivity, leading to single-sided laser output. To account for the excess loss introduced by the external cavity, we replace $P$ in Eq.~\ref{eq:schawlow_townes_linewidth} with $P/(T\eta)$, where $T$ is the on-chip transmission of the external cavity and $\eta$ is the butt-coupling efficiency. The resulting linewidth for hybrid integrated lasers is

\begin{equation}  \label{eq:hybrid_integrated_laser_linewidth}
    \Delta\nu = \frac{\pi h\nu^{3}n_{sp}}{PQ_{l}^{2}}T\eta(1+\alpha_{H}^{2}).
\end{equation}

\subsection{E-DBR laser}

For an E-DBR laser, the external cavity is a waveguide Bragg grating. Its effective cavity length and peak reflectivity are~\cite{tran2019tutorial}

\begin{equation}  \label{eq:grating_Leff_and_reflection}
    \begin{split}
        L_{\rm eff} &= \frac{\tanh(\kappa_{g}l_{g})}{2\kappa_{g}}\\
        R_{g} &= \tanh^{2}(\kappa_{g}l_{g}),
    \end{split}
\end{equation}

where $\kappa_{g}$ is the grating coupling coefficient and $l_{g}$ is the grating length.

In narrow-linewidth E-DBR lasers, the effective cavity length is dominated by the external cavity rather than the gain element. The loaded quality factor of the cold laser cavity then becomes

\begin{equation}  \label{eq:edbr_cavity_Ql}
    Q_{l} = \frac{4\pi n_{g}L_{\rm eff}}{\lambda\ln(R_{g}\eta^{2})},
\end{equation}

where $n_{g}$ is the group index in the external cavity and $\lambda = c/\nu$ is the lasing wavelength. Combining Eqs.~\ref{eq:hybrid_integrated_laser_linewidth}, \ref{eq:grating_Leff_and_reflection}, and \ref{eq:edbr_cavity_Ql} yields an expression for the E-DBR laser linewidth:

\begin{equation}  \label{eq:edbr_laser_linewidth_raw}
    \Delta\nu = \frac{\pi h\nu^{3}n_{sp}}{P}(1-R_{g})\eta(1+\alpha_{H}^{2})\left[\frac{\lambda\ \mathrm{atanh}(\sqrt{R_{g}})\ln(R_{g}\eta^{2})}{2\pi n_{g}l_{g}\sqrt{R_{g}}}\right]^{2},
\end{equation}

where we have used $T = 1 - R_{g}$ for the Bragg grating external cavity. Practical E-DBR lasers typically employ a weak grating ($R_{g} \ll 1$) to balance linewidth and output power~\cite{xiang2019ultra}. This allows the approximation $\mathrm{atanh}(\sqrt{R_{g}}) \approx \sqrt{R_{g}}$, simplifying the linewidth expression to

\begin{equation}  \label{eq:edbr_laser_linewidth}
    \Delta\nu = \frac{\pi h\nu^{3}n_{sp}}{P}(1-R_{g})\eta(1+\alpha_{H}^{2})\left[\frac{\lambda\ln(R_{g}\eta^{2})}{2\pi n_{g}l_{g}}\right]^{2}.
\end{equation}

We next relate the ringdown-limited frequency tuning rate of an E-DBR laser to its intrinsic linewidth. Ringdown occurs in a resonator when high-speed modulation drives the frequency excursion within a photon lifetime beyond the resonance spectral linewidth~\cite{siddharth2025ultrafast}. The ringdown-limited tuning rate is therefore

\begin{equation}  \label{eq:ringdown_tuning_rate}
    \gamma_{\rm rd} = \frac{\Delta\nu_{c}}{\tau_{c}},
\end{equation}

where $\Delta\nu_{c}$ is the 3-dB reflection bandwidth of the external-cavity Bragg grating and $\tau_{c}$ is the corresponding photon lifetime. For weak gratings, $\Delta\nu_{c}$ is approximately half the grating bandwidth, defined as the null-to-null spectral span of the central lobe of its reflection spectrum, and is given by

\begin{equation}  \label{eq:grating_bandwidth}
    \Delta\nu_{c} \approx \frac{c}{2\pi n_{g}l_{g}}\sqrt{\left(\mathrm{atanh}\sqrt{R_{g}}\right)^{2}+\pi^{2}} \approx \frac{c}{2n_{g}l_{g}}.
\end{equation}

The roundtrip photon lifetime in a weak Bragg grating is

\begin{equation}  \label{eq:grating_ph_lifetime}
    \tau_{c} = \frac{2n_{g}L_{\rm eff}}{c} = \frac{n_{g}l_{g}}{c}\frac{\sqrt{R_{g}}}{\mathrm{atanh}(\sqrt{R_{g}})} \approx \frac{n_{g}l_{g}}{c},
\end{equation}

where we have used Eq.~\ref{eq:grating_Leff_and_reflection}.

Substituting Eqs.~\ref{eq:grating_bandwidth} and \ref{eq:grating_ph_lifetime} into Eq.~\ref{eq:ringdown_tuning_rate}, we obtain the ringdown-limited tuning rate for a weak Bragg grating:

\begin{equation}  \label{eq:grating_tuning_rate_limit}
    \gamma_{\rm rd} \approx \frac{1}{2}\left(\frac{c}{n_{g}l_{g}}\right)^{2}.
\end{equation}

Combining the linewidth expression for E-DBR lasers (Eq.~\ref{eq:edbr_laser_linewidth}) with the ringdown-limited tuning rate (Eq.~\ref{eq:grating_tuning_rate_limit}) gives

\begin{equation}  \label{eq:edbr_linewidth_tuning_rate_tradeoff}
    \boxed{\Delta\nu = \frac{h\nu n_{sp}}{2\pi P}\eta(1-R_{g})(1+\alpha_{H}^{2})[\ln(R_{g}\eta^{2})]^{2}\gamma_{\rm rd}}.
\end{equation}

This equation quantitatively captures the fundamental trade-off between the intrinsic linewidth $\Delta\nu$ and the maximum achievable tuning rate $\gamma_{\rm rd}$ in E-DBR lasers.

\subsection{Self-injection-locked laser}

The frequency noise power spectral density of a self-injection-locked laser diode is given by~\cite{ousaid2024low}:

\begin{equation}
    S_{\nu,\mathrm{locked}}(f) = \frac{1+4\left(\frac{f}{\nu}\right)^{2}Q_{r}^{2}}{\left(1+\rho\frac{Q_{r}}{Q_{l}}\right)^{2}+4\left(\frac{f}{\nu}\right)^{2}Q_{r}^{2}}S_{\nu,\mathrm{free}}(f),
\end{equation}

where $S_{\nu,\mathrm{free}}(f)$ is the frequency noise spectrum of the free-running laser diode, $f$ is the offset frequency, $\nu$ is the laser frequency, $Q_{r}$ is the loaded quality factor of the external cavity, $\rho$ is the amplitude reflection coefficient of the external cavity, and $Q_{l}$ is the quality factor of the laser cavity.

Experimentally, the intrinsic linewidth is determined from the white frequency noise level at a specific offset frequency $f'$, using the relations $\Delta\nu_{\mathrm{locked}} = 2\pi S_{\nu,\mathrm{locked}}(f')$ and $\Delta\nu_{\mathrm{free}} = 2\pi S_{\nu,\mathrm{free}}(f')$. This offset frequency is typically much smaller than the spectral linewidth of the external cavity, $f' \ll \Delta\nu_{c} = \nu_{0}/Q_{r}$, an assumption valid for $Q_{r}<10^{7}$. Under these conditions, the linewidth reduction factor due to self-injection locking simplifies to:

\begin{equation}  \label{eq:sil_linewidth_reduction}
    \frac{\Delta\nu_{\mathrm{locked}}}{\Delta\nu_{\mathrm{free}}} \approx \frac{1}{\left(1+\rho\frac{Q_{r}}{Q_{l}}\right)^{2}} = \frac{1}{\left(1+\sqrt{R}\eta\frac{Q_{r}}{Q_{l}}\right)^{2}},
\end{equation}

where $R$ is the power reflection coefficient of the external cavity and $\eta$ is the power coupling efficiency between the laser diode and the external cavity.

Wavelength tuning in a self-injection-locked laser is achieved by tuning its external cavity, which is typically a ring resonator in hybrid integrated lasers. The ring-down-limited tuning rate is:

\begin{equation}  \label{eq:sil_tuning_rate}
    \gamma_{\rm rd} = \frac{\Delta\nu_{c}}{\tau_{c}} = 2\pi\Delta\nu_{c}^{2} = 2\pi\left(\frac{\nu}{Q_{r}}\right)^{2},
\end{equation}

where we have used the ring photon lifetime $\tau_{c} = (2\pi\Delta\nu_{c})^{-1}$. By combining Eqs.~\ref{eq:sil_linewidth_reduction} and \ref{eq:sil_tuning_rate}, we obtain the trade-off between linewidth and maximum tuning rate:

\begin{equation}  \label{eq:sil_linewidth_tuning_rate_tradeoff_raw}
    \Delta\nu_{\mathrm{locked}} = \frac{Q_{l}^{2}}{\left(Q_{l}+\eta\nu\sqrt{\frac{2\pi R}{\gamma_{\rm rd}}}\right)^{2}}\Delta\nu_{\mathrm{free}}.
\end{equation}

The free-running linewidth $\Delta\nu_{\mathrm{free}}$ follows the Schawlow-Townes formula (Eq.~\ref{eq:hybrid_integrated_laser_linewidth}). Substituting this expression into Eq.~\ref{eq:sil_linewidth_tuning_rate_tradeoff_raw} yields the final formulation for the linewidth-agility trade-off in SIL lasers:

\begin{equation}  \label{eq:sil_linewidth_tuning_rate_tradeoff}
    \boxed{\Delta\nu_{\mathrm{locked}} = \frac{\pi h\nu^{3}n_{sp}}{P\left(Q_{l}+\eta\nu\sqrt{\frac{2\pi R}{\gamma_{\rm rd}}}\right)^{2}}T\eta(1+\alpha_{H}^{2}).}
\end{equation}

\subsection{RE-DBR laser}

The frequency noise characteristics of RE-DBR lasers were theoretically analyzed in Ref.~\cite{yu2026resonator}. Taking into account the wavelength selectivity of the external cavity, the intrinsic linewidth is given by

\begin{equation}  \label{eq:redbr_linewidth_raw}
    \Delta\nu = \Delta\nu_{0}\left(\frac{n_{g,a}l_{a}\pi}{n_{g}l_{r}\mathcal{F}}\right)^{2},
\end{equation}

where $\Delta\nu_{0}$ denotes the linewidth of a Fabry-Pérot laser diode with a front mirror reflectivity equal to the peak reflectivity of the RE-DBR, $n_{g,a}$ and $n_{g}$ are the group indices of the gain element and the external cavity, respectively, $l_{a}$ is the length of the gain chip, $l_{r}$ is the circumference of the RE-DBR, and $\mathcal{F}$ represents the finesse of the RE-DBR resonator. This expression is derived by combining the general Lorentzian linewidth formula for external-cavity lasers with an effective cavity length obtained from the coupled-mode theory of the grating-assisted ring resonator.

The Fabry-Pérot laser linewidth $\Delta\nu_{0}$ is given by Eq.~\ref{eq:hybrid_integrated_laser_linewidth}, with the cavity quality factor defined as

\begin{equation}  \label{eq:fp_q_factor}
    Q_{l} = \frac{4\pi n_{g,a}l_{a}}{\lambda\ln(R_{r}\eta^{2})},
\end{equation}

where $R_{r}$ is the peak reflectivity of the RE-DBR external cavity. The resonator finesse is related to the resonator quality factor $Q_{r}$ by

\begin{equation}  \label{eq:redbr_finesse}
    \mathcal{F} = \frac{\lambda Q_{r}}{n_{g}l_{r}}.
\end{equation}

Combining these equations yields the intrinsic linewidth for a RE-DBR laser:

\begin{equation}  \label{eq:redbr_linewidth}
    \Delta\nu = \frac{\pi h\nu^{3}n_{sp}}{16PQ_{r}^{2}}T\eta(1+\alpha_{H}^{2})[\ln(R_{r}\eta^{2})]^{2},
\end{equation}

where $Q_{r}$, $R$, and $T$ are the loaded quality factor, peak reflectivity, and transmittance at the peak-reflection wavelength of the RE-DBR external cavity, respectively; all other parameters are defined as for E-DBR lasers. Finally, combining this linewidth expression (Eq.~\ref{eq:redbr_linewidth}) with the ring-down-limited tuning rate for a ring resonator (Eq.~\ref{eq:sil_tuning_rate}) gives the linewidth-agility trade-off in a RE-DBR laser:

\begin{equation}  \label{eq:redbr_tradeoff}
    \boxed{\Delta\nu = \frac{h\nu n_{sp}}{32P}T\eta(1+\alpha_{H}^{2})[\ln(R_{r}\eta^{2})]^{2}\gamma_{\rm rd}.}
\end{equation}

\subsection{Numerical simulations}

\begin{figure}[t!]
    \centering
    \includegraphics{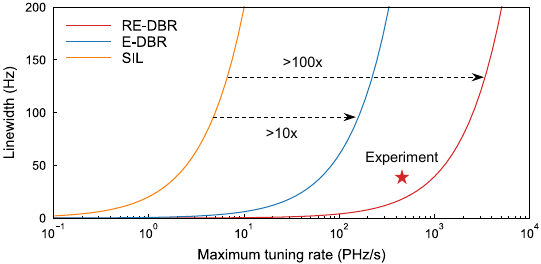}
    \captionsetup{singlelinecheck=off, justification = RaggedRight}
    \caption{
        \textbf{Tradeoff between ringdown-limited tuning rate and intrinsic linewidth.} The red star indicates our experimental configuration of the RE-DBR laser. The RE-DBR laser can achieve a higher tuning rate than E-DBR and SIL lasers at the same linewidth.
    }
    \label{fig:linewidth_tuning_rate_tradeoff}
\end{figure}

To quantitatively compare the trade-off between linewidth and frequency agility in E-DBR, SIL, and RE-DBR lasers, we plot the theoretical intrinsic linewidth against the ringdown-limited tuning rate using Eqs.~\ref{eq:edbr_linewidth_tuning_rate_tradeoff}, \ref{eq:sil_linewidth_tuning_rate_tradeoff}, and \ref{eq:redbr_tradeoff}. The result is shown in Fig.~\ref{fig:linewidth_tuning_rate_tradeoff}, and the corresponding parameters are summarized in Table~\ref{tab:tradeoff_calc_parameters}. For a fair comparison, all lasers are assumed to operate at 1550 nm with an output power of 1 mW. The loaded cavity quality factor of the laser diode, $Q_l$, is set to $2.5\times10^{3}$ following Ref.~\cite{ousaid2024low}. The reflectivity of the microring resonator induced by Rayleigh scattering is set to 3\%~\cite{snigirev2023ultrafast,xiang20233d}, whereas the peak reflectivities of both the Bragg grating and the RE-DBR external cavity are assumed to be 40\%~\cite{yu2026resonator,xiang2019ultra}.

\begin{table}[h]
    \centering
    \renewcommand{\arraystretch}{1.25}
    \setlength{\tabcolsep}{6pt}
    \small
    
    \begin{tabular}{|c|c|c|c|}
    \hline
    \multicolumn{1}{|c|}{\textbf{Symbol}} &
    \textbf{Value} &
    \textbf{Unit} &
    \textbf{Definition} \\
    \hline

    $h$ & $6.63\times10^{-34}$ & kg~m\textsuperscript{2}\,s\textsuperscript{-1} & Planck's constant \\
    $n_{sp}$ & 3 & 1 & Population inversion factor \\
    $\alpha_{H}$ & 2 & 1 & Amplitude-phase coupling factor \\
    $\nu$ & 193.4 & THz & Laser emission frequency \\
    $\eta$ & $-5$ & dB & Butt-coupling efficiency \\
    $P$ & 1 & mW & Laser output power \\
    $Q_{l}$ & $2.5\times 10^{3}$ & 1 & Cavity loaded $Q$ of the free-running laser diode \\
    $T$ & 0.2 & 1 & Transmission of external cavity for SIL or RE-DBR laser \\
    $R$ & 0.03 & 1 & Reflectivity due to Rayleigh scattering in the microring resonator \\
    $R_{g}$ & 0.4 & 1 & Peak reflectivity of the Bragg grating \\
    $R_{r}$ & 0.4 & 1 & Peak reflectivity of the RE-DBR external cavity \\
    $\gamma_{\rm rd}$ & $10^{-1}-10^{4}$ & PHz & Ringdown-limited tuning rate \\
    
    \hline
    \end{tabular}
    
    \caption{\textbf{Parameters used in the simulations of laser linewidth-agility trade-off.}}
    \label{tab:tradeoff_calc_parameters}

\end{table}

The simulations show that in all three laser architectures, the achievable tuning rate decreases as the linewidth narrows. However, at a fixed linewidth, the RE-DBR laser supports a substantially higher tuning rate than both the E-DBR and self-injection-locked lasers, thereby mitigating the linewidth-agility trade-off that constrains these conventional architectures. This improvement arises from two key properties of the RE-DBR reflector: its feedback strength and its temporal response. First, the RE-DBR is designed to provide strong peak reflectivity, which narrows the linewidth while maintaining a relatively low external-cavity quality factor. This reduces the limitation that finite photon lifetime and cavity ringdown impose on frequency agility. Second, the RE-DBR reflector has an impulse response with an exponentially decaying reflected intensity. Consequently, during frequency modulation, optical feedback is dominated by spectral components near the instantaneous resonant frequency, while feedback from earlier frequencies decays rapidly. This temporal weighting suppresses the adverse effects of cavity memory on laser dynamics and increases the ringdown-limited tuning rate. In contrast, a waveguide Bragg grating produces an approximately constant reflected intensity over a finite temporal window, making E-DBR lasers more susceptible to external-cavity ringdown and thus limiting their ringdown-limited tuning rate.

\section{Analysis of laser chirp linearity} \label{SI_linearity_limited_by_bandwidth}

The frequency modulation linearity of a laser degrades as the modulation frequency approaches the modulation bandwidth. In this section, we present a theoretical analysis of this bandwidth-limited chirp linearity in a RE-DBR laser. Our calculations indicate that the experimentally observed increase in modulation nonlinearity with frequency arises primarily from the finite modulation bandwidth.

To model the temporal evolution of the external cavity laser, we employ temporal coupled mode theory. The coupled mode equations for the RE-DBR external cavity are~\cite{jin2022ultra}

\begin{equation}  \label{eq:temporal_coupled_mode_eq}
    \begin{split}
        \frac{d}{dt}A_{cw} &= \left(i\omega_{0}-\frac{1}{2\tau}\right)A_{cw} + i\beta A_{ccw} + b S\\
        \frac{d}{dt}A_{ccw} &= i\beta A_{cw} + \left(i\omega_{0}-\frac{1}{2\tau}\right)A_{ccw},
    \end{split}
\end{equation}

where $A_{cw}$ ($A_{ccw}$) represents the complex amplitude of the clockwise (counter-clockwise) mode, $\omega_{0}$ is the angular resonant frequency, $\tau$ is the photon lifetime, $\beta$ is the grating coupling strength, and $b$ is the coupling strength between the resonator and the bus waveguide.

To analyze the frequency modulation response, we assume a time-dependent resonant frequency $\omega_{0}(t) = \omega_{dc} + \Delta\omega(t)$, where $\omega_{dc}$ and $\Delta\omega(t)$ are the DC and AC components, respectively. To simplify the formulation, we adopt the substitutions $A_{cw(ccw)} = A_{cw(ccw)}'\exp(i\omega_{dc}t)$ and $S = S'\exp(i\omega_{dc}t)$, yielding

\begin{equation}  \label{eq:temporal_coupled_mode_eq_rot_frame}
    \begin{split}
        \frac{d}{dt}A_{cw}' &= \left(i\Delta\omega(t) - \frac{1}{2\tau}\right)A_{cw}' + i\beta A_{ccw}' + bS'\\
        \frac{d}{dt}A_{ccw}' &= i\beta A_{cw}' + \left(i\Delta\omega(t) - \frac{1}{2\tau}\right)A_{ccw}'.
    \end{split}
\end{equation}

Next, we assume that the input light coincides with the ring's static resonant frequency $\omega_{dc}$, resulting in a time-independent $S'$. This assumption is valid for the RE-DBR laser, which spontaneously operates at the external-cavity resonant frequency. We then split the mode amplitudes into a steady-state component and a small-signal perturbation:

\begin{equation}  \label{eq:assumption_small_signal_analysis}
    \begin{split}
        A_{cw}' &= A_{cw,0}' + \delta A_{cw}'(t)\\
        A_{ccw}' &= A_{ccw,0}' + \delta A_{ccw}'(t),
    \end{split}
\end{equation}

where $A_{cw,0}'$ and $A_{ccw,0}'$ represent the steady-state solutions, which satisfy

\begin{equation}  \label{eq:steady_state_solution}
    \begin{split}
        0 &= -\frac{1}{2\tau}A_{cw,0}' + i\beta A_{ccw,0}' + bS'\\
        0 &= i\beta A_{cw,0}' - \frac{1}{2\tau}A_{ccw,0}'.
    \end{split}
\end{equation}

Combining these equations and neglecting higher-order small terms leads to the following small-signal perturbation equations:

\begin{equation}  \label{eq:small_signal_perturbation}
    \begin{split}
        \frac{d}{dt}\delta A_{cw}'(t) &= i\Delta\omega(t)A_{cw,0}' - \frac{1}{2\tau}\delta A_{cw}'(t) + ib\delta A_{ccw}'(t)\\
        \frac{d}{dt}\delta A_{ccw}'(t) &= i\Delta\omega(t)A_{ccw,0}' - \frac{1}{2\tau}\delta A_{ccw}'(t) + ib\delta A_{cw}'(t).
    \end{split}
\end{equation}

Note that both \(A_{cw}'\) and \(A_{ccw}'\) depend on the frequency detuning between the laser and the cavity resonance. Their magnitudes reach local maxima at zero detuning, where the first-order derivative of the amplitude with respect to detuning must vanish. Consequently, small changes in the mode amplitudes, \(\delta A_{cw}'\) and \(\delta A_{ccw}'\), induced by external-cavity frequency modulation, can be expressed as pure phase perturbations:

\begin{equation}  \label{eq:small_signal_perturbation_as_phase_shift}
    \begin{split}
        \delta A_{cw}'(t) &= i\,\delta\phi_{cw}(t)\,A_{cw,0}' \\
        \delta A_{ccw}'(t) &= i\,\delta\phi_{ccw}(t)\,A_{ccw,0}'.
    \end{split}
\end{equation}

Here, the unperturbed amplitudes are written as \(A_{cw,0}' = |A_{cw,0}'|\exp(i\phi_{cw,0})\) and \(A_{ccw,0}' = |A_{ccw,0}'|\exp(i\phi_{ccw,0})\), with \(\delta\phi_{cw}\) and \(\delta\phi_{ccw}\) representing small phase deviations.

Substituting Eq.~\ref{eq:small_signal_perturbation_as_phase_shift} into Eqs.~\ref{eq:small_signal_perturbation} and using the steady-state relations from Eqs.~\ref{eq:steady_state_solution} to simplify the result, we obtain a set of coupled differential equations for \(\delta\phi_{cw}(t)\) and \(\delta\phi_{ccw}(t)\):

\begin{equation} \label{eq:delta_phi}
    \begin{split}
        \frac{d}{dt}\delta\phi_{cw} &= \Delta\omega(t) - \frac{1}{2\tau}\delta\phi_{cw} - 2\tau b^{2}\,\delta\phi_{ccw} \\
        \frac{d}{dt}\delta\phi_{ccw} &= \Delta\omega(t) - \frac{1}{2\tau}\delta\phi_{ccw} + \frac{1}{2\tau}\delta\phi_{cw}.
    \end{split}
\end{equation}

The phase shift \(\delta\phi_{cw}\) corresponds directly to a shift in the laser frequency. To determine this frequency shift, we differentiate the resonant condition of the lasing mode,

\begin{equation} \label{eq:resonant_condition}
    2\frac{2\pi n_{r}L_{\rm eff}}{\lambda} + 2\frac{2\pi n_{a}l_{a}}{\lambda} + 2\varphi = 2N\pi,
\end{equation}

where \(n_{r}\) and \(n_{a}\) are the effective mode indices in the RE-DBR and the gain element, respectively, \(L_{\rm eff}\) is the effective cavity length of the RE-DBR, \(l_{a}\) is the length of the gain element, \(N\) is an integer representing the longitudinal mode index, and \(\varphi\) denotes the phase shift induced by frequency modulation. Since \(L_{\rm eff}\) is maximized at zero detuning, its first-order derivative with respect to detuning vanishes. Differentiating Eq.~\ref{eq:resonant_condition} then yields the following expression for the resonant frequency shift due to modulation:

\begin{equation} \label{eq:freq_shift}
    \delta f = -\frac{c}{4\pi(n_{a}l_{a} + n_{r}L_{\rm eff})}\,\delta\phi_{cw} \equiv K\,\delta\phi_{cw}.
\end{equation}

Combining Eqs.~\ref{eq:delta_phi} and \ref{eq:freq_shift} to eliminate \(\delta\phi_{cw}\) and \(\delta\phi_{ccw}\) produces the time-domain equation governing the frequency modulation response of the RE-DBR laser:

\begin{equation} \label{eq:redbr_response_time_domain}
    \left(\frac{d^{2}}{dt^{2}} + \frac{1}{\tau}\frac{d}{dt} + \frac{1}{4\tau^{2}} + b^{2}\right)\delta f = K\left(\frac{d}{dt} + \frac{1}{2\tau} - 2\tau b^{2}\right)\Delta\omega(t).
\end{equation}

Taking the Laplace transform of Eq.~\ref{eq:redbr_response_time_domain} gives the transfer function from the resonator frequency change \(\Delta\omega\) to the laser frequency shift \(\delta f\):

\begin{equation} \label{eq:redbr_transfer_func_raw}
    H(\omega) = K\,\frac{i\omega - 2\tau\!\left(b^{2} - \frac{1}{4\tau^{2}}\right)}{\left(i\omega + \frac{1}{2\tau}\right)^{2} + b^{2}},
\end{equation}

where \(\omega\) in this context is the angular modulation frequency. In the weak grating limit (\(b \ll 1/2\tau\)), the transfer function simplifies to

\begin{equation} \label{eq:redbr_transfer_func}
    \boxed{H(\omega) \approx \frac{2\tau K}{1 + i\,2\tau\omega}.}
\end{equation}

This expression shows that the frequency modulation response of a RE-DBR laser is analogous to that of a first-order low-pass filter, with a 3-dB bandwidth determined by the photon lifetime in the resonator: \(\omega_{\rm 3dB} = 1/2\tau\).

We now analyze the laser modulation nonlinearity arising from the finite modulation bandwidth. For triangular-wave modulation, the time variation of the resonator's angular resonant frequency is given by  

\begin{equation}  \label{eq:tri_wave_decomp}
    \Delta\omega(t) = 4\frac{2A}{\pi^{2}}\sum^{\infty}_{n=1,3,5,\dots}\frac{(-1)^{\frac{n-1}{2}}}{n^{2}}\sin(n\omega_{m}t),
\end{equation}  

where \(2A\) is the peak-to-peak amplitude of \(\Delta\omega(t)\) and \(\omega_{m}\) is the triangular-wave modulation frequency.  

Using the frequency-modulation transfer function in Eq.~\ref{eq:redbr_transfer_func}, the corresponding laser frequency excursion can be derived as  

\begin{align}  \label{eq:laser_freq_excursion}
    \delta f(t) &= \frac{8A}{\pi^{2}}\sum_{n=1,3,5,\dots}^{\infty}\frac{(-1)^{\frac{n-1}{2}}}{n^{2}}H(n\omega_{m})\sin(n\omega_{m}t)\nonumber\\
    &= A\frac{16\tau K}{\pi^{2}}\sum_{n=1,3,5,\dots}^{\infty}\frac{(-1)^{\frac{n-1}{2}}}{n^{2}}\frac{1}{1+i2\tau n\omega_{m}}\sin(n\omega_{m}t).
\end{align}  

Note that this expression converges to a perfect triangular-wave chirp in the limit \(\tau\rightarrow 0\). The ideal laser frequency excursion is then  

\begin{equation}  \label{eq:laser_freq_excursion_ideal}
    \delta f_{0}(t) = 4\frac{4\tau KA}{\pi^{2}}\sum_{n=1,3,5,\dots}^{\infty}\frac{(-1)^{\frac{n-1}{2}}}{n^{2}}\sin(n\omega_{m}t).
\end{equation}  

Comparing Eq.~\ref{eq:laser_freq_excursion} and Eq.~\ref{eq:laser_freq_excursion_ideal} yields the frequency error as a function of time:  

\begin{equation}  \label{eq:freq_error}
    e(t) = \delta f(t) - \delta f_{0}(t) = -iA\frac{8\tau K}{\pi^{2}}\sum_{n=\pm1,\pm3,\dots}^{\infty}\frac{(-1)^{\frac{n-1}{2}}}{n^{2}}\left[\frac{1}{1+i\ {\rm sgn}(n)2\tau n\omega_{m}}-1\right]e^{in\omega_{m}t}.
\end{equation}  

The root-mean-square (RMS) frequency error can be calculated using Parseval's theorem, yielding  

\begin{equation}  \label{eq:freq_error_rms}
    e_{\rm RMS} = Ar\frac{16\sqrt{2}\tau K}{\pi^{2}}\sqrt{\sum_{n=1,3,5,\dots}^{\infty}\frac{1}{n^{2}(1+4n^{2}r^{2})}},
\end{equation}  

where \(r = \tau\omega_{m} = f_{m}/\Delta\nu_{c}\) is the normalized modulation frequency, \(f_{m}\) the modulation frequency, and \(\Delta\nu_{c}\) the resonance linewidth. The RMS modulation nonlinearity \(\zeta\) (also called chirp nonlinearity) is defined as the RMS frequency error \(e_{\rm RMS}\) divided by the peak-to-peak laser frequency excursion \(\mathrm{BW} = 4\tau KA\):  

\begin{equation}  \label{eq:rms_mod_nonlinearity}
    \boxed{\zeta = \frac{e_{\rm RMS}}{\mathrm{BW}} = \frac{4\sqrt{2}r}{\pi^{2}}\sqrt{\sum_{n=1,3,5,\dots}^{\infty}\frac{1}{n^{2}(1+4n^{2}r^{2})}}.}
\end{equation}  

\begin{figure}[t!]
    \centering
    \includegraphics{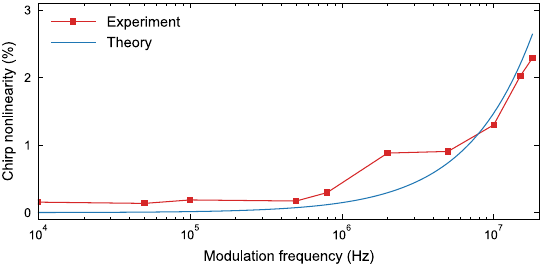}
    \captionsetup{singlelinecheck=off, justification = RaggedRight}
    \caption{
        \textbf{Comparison of theoretical and measured chirp nonlinearity of the RE-DBR laser.} The theoretical calculation of chirp nonlinearity assumes a resonance linewidth of 389 MHz. 
    }
    \label{fig:chirp_nonlinearity}
\end{figure}

We use this equation to calculate the chirp nonlinearity of the RE-DBR laser and compare the results with experimentally measured values, as shown in Fig.~\ref{fig:chirp_nonlinearity}. The calculation assumes a spectral linewidth of 389 MHz for the RE-DBR resonance; the experimental methods are detailed in the main text. Theory and experiment show good agreement. Since our chirp-nonlinearity model accounts only for the bandwidth limitation imposed by the finite photon lifetime, this agreement indicates that the chirp nonlinearity of our RE-DBR laser originates primarily from the finite external-cavity photon lifetime.  

\section{Reflection characteristics of microring resonators under high-speed modulation}
\label{SI_dynamic_resonance}

Self-injection-locked lasers commonly employ high-\(Q\) microring resonators as external cavities to provide narrowband feedback, thereby achieving narrow linewidths. However, as we show below, when such resonators undergo high-speed modulation, their reflection characteristics can change significantly, degrading the laser coherence and potentially undermining single-mode operation.

To analyze this behavior, we establish a theoretical model for the reflection characteristics of microring resonators under high-speed modulation. We consider a critically coupled ring resonator with internal scatterers subjected to linear frequency modulation, where the instantaneous resonant frequency is \(\omega_{0}(t) = \omega_{dc} + \gamma t\). The time evolution of the optical field inside the resonator is then governed by the following temporal coupled-mode equations (recall Eq.~\ref{eq:temporal_coupled_mode_eq_rot_frame}):

\begin{equation}
    \label{eq:ring_temporal_coupled_eq}
    \begin{split}
        \frac{d}{dt}A_{cw}' &= \left(i\Delta\omega(t)-\frac{1}{2\tau}\right)A_{cw}' + i\beta A_{ccw}' + \sqrt{\frac{1}{2\tau}}s_{0}, \\
        \frac{d}{dt}A_{ccw}' &= i\beta A_{cw}' + \left(i\Delta\omega(t)-\frac{1}{2\tau}\right)A_{ccw}'.
    \end{split}
\end{equation}

Here, the field amplitudes are normalized such that \(|A_{cw}'|^{2}\) and \(|A_{ccw}'|^{2}\) represent the energy in the corresponding modes, and \(|s_{0}|^{2}\) denotes the input power on the bus waveguide. The instantaneous detuning is defined as \(\Delta\omega(t) = \omega_{0}(t) - \omega\), where \(\omega\) is the angular pump frequency; all other parameters follow the definitions in Eq.~\ref{eq:temporal_coupled_mode_eq_rot_frame}. Without loss of generality, we assume the backscattering strength coefficient \(\beta\) to be real and positive. The transmitted and reflected fields on the bus waveguide are then given by

\begin{equation}
    \label{eq:bus_wg_field_amp}
    \begin{split}
        s_{\rm out} &= s_{0} - \frac{1}{2\tau}A_{cw}', \\
        s_{\rm ref} &= -\frac{1}{2\tau}A_{ccw}'.
    \end{split}
\end{equation}

To solve Eq.~\ref{eq:ring_temporal_coupled_eq}, we introduce the following supermode amplitudes:

\begin{equation}
    \label{eq:supermode_amp}
    \begin{split}
        A &= \frac{A'_{cw} + A'_{ccw}}{\sqrt{2}}, \\
        B &= \frac{A'_{cw} - A'_{ccw}}{\sqrt{2}}.
    \end{split}
\end{equation}

Substituting Eq.~\ref{eq:supermode_amp} into Eq.~\ref{eq:ring_temporal_coupled_eq} diagonalizes the system, yielding two decoupled equations:

\begin{equation}
    \label{eq:ring_temporal_coupled_eq_reduced}
    \begin{split}
        \frac{dA}{dt} &= \left(i(\Delta\omega(t)+\beta)-\frac{1}{2\tau}\right)A + \frac{1}{2\sqrt{\tau}}s_{0}, \\
        \frac{dB}{dt} &= \left(i(\Delta\omega(t)-\beta)-\frac{1}{2\tau}\right)B + \frac{1}{2\sqrt{\tau}}s_{0}.
    \end{split}
\end{equation}

The solution to these first-order ordinary differential equations is

\begin{equation}
    \label{eq:ring_temporal_coupled_eq_sol}
    \begin{split}
        A(\Delta\omega) &= \frac{s_{0}}{2}\sqrt{\frac{\pi}{2i\tau\gamma}}\, w\!\left(\frac{1}{\sqrt{2i\gamma}}\left(\Delta\omega+\beta+\frac{i}{2\tau}\right)\right), \\
        B(\Delta\omega) &= \frac{s_{0}}{2}\sqrt{\frac{\pi}{2i\tau\gamma}}\, w\!\left(\frac{1}{\sqrt{2i\gamma}}\left(\Delta\omega-\beta+\frac{i}{2\tau}\right)\right),
    \end{split}
\end{equation}

where \(w(z) = \exp(-z^{2})\,\mathrm{erfc}(-iz)\) is the Faddeeva function. Because the detuning depends linearly on time, i.e., \(\Delta\omega = \omega_{dc} + \gamma t - \omega\), we can express the solution directly as a function of the instantaneous detuning \(\Delta\omega\). Combining Eqs.~\ref{eq:ring_temporal_coupled_eq_sol}, \ref{eq:supermode_amp}, and \ref{eq:bus_wg_field_amp}, we obtain the instantaneous reflection spectrum of the ring resonator under frequency modulation:

\begin{equation}
    \label{eq:dynamic_reflection}
    \boxed{
        R(\Delta\omega) = \left|\frac{s_{\rm ref}}{s_{0}}\right|^{2}
        = \left|\frac{1}{4\tau}\sqrt{\frac{\pi}{2\gamma}}
        \left[
            w\!\left(\frac{1}{\sqrt{2i\gamma}}\left(\Delta\omega+\beta+\frac{i}{2\tau}\right)\right)
            - w\!\left(\frac{1}{\sqrt{2i\gamma}}\left(\Delta\omega-\beta+\frac{i}{2\tau}\right)\right)
        \right]
        \right|^{2}.
    }
\end{equation}

\begin{figure}[t!]
    \centering
    \includegraphics{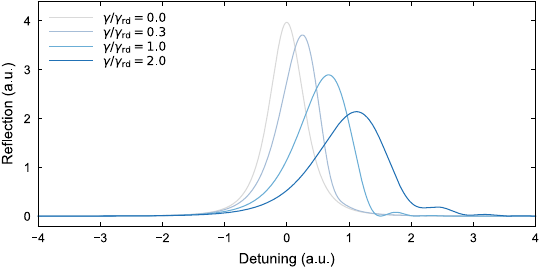}
    \captionsetup{singlelinecheck=off, justification=RaggedRight}
    \caption{
        \textbf{Instantaneous reflection spectrum of a ring resonator under frequency modulation.}
        The spectrum broadens at high tuning rates \(\gamma\), degrading laser coherence and the stability of single-mode operation.
    }
    \label{fig:ring_dynamic_reflection}
\end{figure}

Using this equation, we numerically simulate the reflection characteristics of a ring resonator, as shown in Fig.~\ref{fig:ring_dynamic_reflection}. We fix the backscattering strength at \(\beta = 0.03\tau^{-1}\) and vary the tuning rate \(\gamma\). The results reveal that the reflection spectrum broadens significantly when \(\gamma\) surpasses the ringdown-limited tuning rate \(\gamma_{\rm rd} = 2\pi\Delta\nu_{c}^{2} = (2\pi\tau^{2})^{-1}\). This spectral broadening can degrade laser coherence and even induce mode hops, thereby restricting stable single-mode operation at high tuning rates. Consequently, a lower cavity \(Q\) (i.e., a larger \(\Delta\nu_{c}\)) is desirable for maintaining single-mode, narrow-linewidth operation during high-speed modulation.

\section{Parameter optimization of RE-DBR external cavity} \label{SI_param_optimize}

\begin{figure}[t!]
    \centering
    \includegraphics{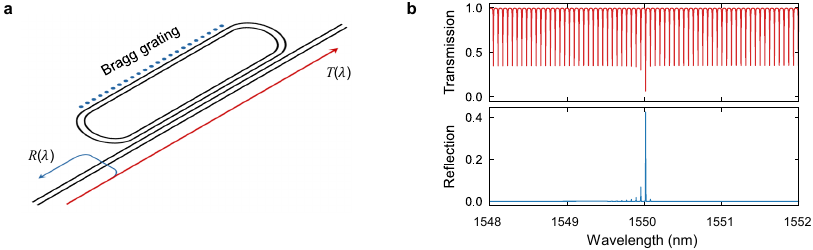}
    \captionsetup{singlelinecheck=off, justification = RaggedRight}
    \caption{
        \textbf{a}, Schematic of the RE-DBR external cavity. \textbf{b}, Simulated transmission and reflection spectra for a RE-DBR consisting of an 8-mm-long grating and a 19-mm-long racetrack resonator. Other key simulation parameters include a resonator coupling coefficient $\kappa_{c}l_{c} = 0.15\pi$, a grating coupling coefficient $\kappa_{g} = 0.1$ cm\textsuperscript{-1}, and an intrinsic loss of 0.15 dB\,cm\textsuperscript{-1}. The waveguide effective index $n_{\rm eff}$ is set to 1.84 for both the resonator and the bus waveguide.
    }
    \label{fig:coupled_mode_theory}
\end{figure}

We analyze the transmittance and reflectance characteristics of a RE-DBR external cavity using coupled mode theory~\cite{yu2026resonator,kang2009microring,kang2010engineering}. This semi-analytical approach is computationally efficient and facilitates the optimization of design parameters. Specifically, we consider a racetrack resonator evanescently coupled to a bus waveguide via a directional coupler, with a periodic array of holes forming a Bragg grating along one straight section, as depicted in Fig.~\ref{fig:coupled_mode_theory}a.

The transmission behavior of this composite structure is governed by several design parameters: the resonator coupling strength $\kappa_{c}$, the grating coupling strength $\kappa_{g}$, the grating period $\Lambda$ and order $N$, the lengths of the Bragg grating ($l_{g}$), the resonator coupler ($l_{c}$), and the resonator itself ($l_{r}$). Additional factors include the waveguide intrinsic loss $\rho$, and the propagation constants in the bus waveguide $\beta_{\rm bus}$ and in the resonator $\beta_{\rm ring}$. Based on these parameters, the amplitude transmission and reflection coefficients of the RE-DBR are expressed as

\begin{equation}  \label{eq:coupled_mode_theory}
    \begin{split}
        t &= \tau_{1} + \tau_{2}\tau_{2}^{*}t_{0}\frac{\tau_{1}^{*}t_{0}-|t_{g}|}{1-2\tau_{1}^{*}|t_{g}|t_{0}+(\tau_{1}^{*})^{2}t_{0}^{2}}\\
        r &= i\frac{\tau_{2}\tau_{2}^{*}|r_{g}|t_{0}}{1-2\tau_{1}^{*}|t_{g}|t_{0}+(\tau_{1}^{*})^{2}t_{0}^{2}}.
    \end{split}
\end{equation}

Here, $\tau_{1} = [\cos(ql_{c})+i\frac{\delta}{q}\sin(ql_{c})]e^{-i\delta l_{c}}$ and $\tau_{2} = -i\frac{\kappa_{c}}{q}\sin(ql_{c})e^{-i\delta l_{c}}$ represent the self-coupling and cross-coupling coefficients of the directional coupler, respectively, with the effective coupling strength $q = \sqrt{\kappa_{c}^{2}+\delta^{2}}$ and the propagation constant mismatch $\delta = (\beta_{\rm ring}-\beta_{\rm bus})/2$. The roundtrip amplitude transmission coefficient is $t_{0} = \exp(-\rho l_{r}/2 - i\beta_{\rm ring}l_{r})$. The amplitude transmission and reflection coefficients of the Bragg grating, $t_{g}$ and $r_{g}$, are given by~\cite{okamoto2021fundamentals}

\begin{equation}  \label{eq:grating_coefficients}
    \begin{split}
        t_{g} &= \frac{\rho_{g}}{\rho_{g}\cos(\rho_{g}l_{g}) + i\phi\sin(\rho_{g}l_{g})}\exp(i\phi l_{g})\\
        r_{g} &= -\frac{i\kappa_{g}\sin(\rho_{g}l_{g})}{\rho_{g}\cos(\rho_{g}l_{g})+i\phi\sin(\rho_{g}l_{g})},
    \end{split}
\end{equation}

with $\phi = \beta_{\rm ring} - N\pi/\Lambda$ and $\rho_{g} = \sqrt{\phi^{2}-\kappa_{g}^{2}}$.

Once the propagation constants $\beta_{\rm ring}$ and $\beta_{\rm bus}$, along with the coupling coefficients $\kappa_{c}$ and $\kappa_{g}$, are obtained from numerical calculations (e.g., finite-element simulations), the transmittance and reflectance spectra $T(\lambda) = |t|^{2}$ and $R(\lambda) = |r|^{2}$ can be calculated using Eq.~\ref{eq:coupled_mode_theory}. Notably, when the Bragg wavelength of the grating coincides with a resonator mode, the RE-DBR exhibits a single-peaked reflectance spectrum. Figure~\ref{fig:coupled_mode_theory} shows the simulated spectra for this configuration, where the Bragg wavelength of an 8-mm-long grating is aligned with a resonance of a 19-mm-long racetrack resonator, yielding a prominent reflection peak.

\begin{figure}[t!]
    \centering
    \includegraphics{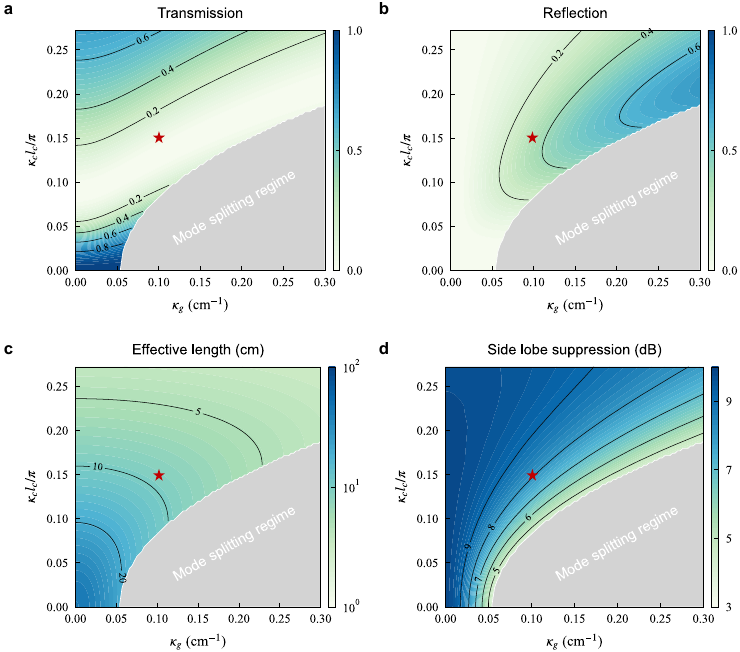}
    \captionsetup{singlelinecheck=off, justification = RaggedRight}
    \caption{
        \textbf{Simulated RE-DBR external-cavity characteristics.} \textbf{a}, Transmission at the wavelength of maximum reflectivity. \textbf{b}, Maximum reflectivity. \textbf{c}, Effective cavity length at the wavelength of maximum reflectivity. \textbf{d}, Side-lobe reflection suppression, defined as the ratio of the maximum reflectivity to the second-highest reflectivity. The gray region indicates parameter sets where the highest reflection peak splits into two peaks, potentially leading to multimode lasing. The red star marks the parameter configuration used in our experimental demonstration of the RE-DBR laser.
    }
    \label{fig:param_optimize}
\end{figure}

To realize a single-mode, narrow-linewidth laser, the RE-DBR external cavity must provide strong, single-peaked reflection combined with a long effective cavity length. High transmittance is also desirable to achieve sufficient output power for practical applications. The effective cavity length can be calculated as $L_{\rm eff} = i\frac{c}{2n_{g}r}\frac{\partial r}{\partial\omega}$, where $r = r(\omega)$ is the reflectivity and $n_{g}$ is the group index in the resonator. To identify the optimal parameter set for narrow-linewidth operation, we sweep the coupling strengths of the grating ($\kappa_{g}$) and the directional coupler ($\kappa_{c}$), both of which can be experimentally tuned by adjusting the separation between the resonator and the bus waveguide or the Bragg grating. Figure~\ref{fig:param_optimize} shows the simulated maximum reflectivity, transmittance and effective cavity length at the wavelength of peak reflectivity, and the side-lobe reflection suppression ratio as functions of these parameters. All other simulation parameters match those used in Fig.~\ref{fig:coupled_mode_theory}b. Based on these results, we select the parameter configuration $\kappa_{g} = 0.1$ cm\textsuperscript{-1} and $\kappa_{c}l_{c} = 0.15\pi$ (marked by the red star) for our experimental demonstration. This configuration balances high transmission, high reflectivity, long effective cavity length, and strong side-lobe suppression, thereby enabling stable single-mode, narrow-linewidth lasing with moderate output power.

\section{Design of low-loss second-order Bragg gratings} \label{SI_2nd_order_grating}

\begin{figure}[t!]
    \centering
    \includegraphics{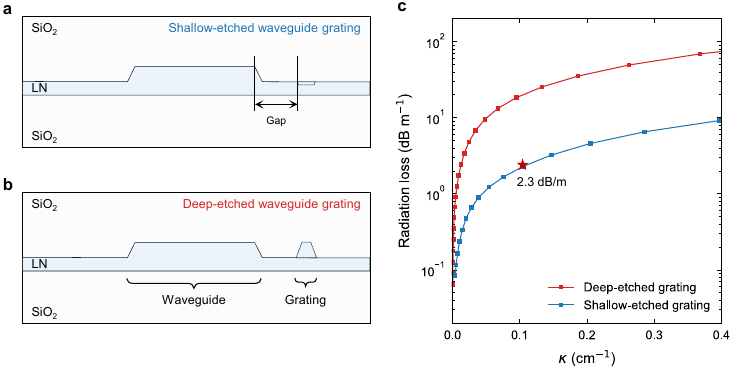}
    \captionsetup{singlelinecheck=off, justification = RaggedRight}
    \caption{
        \textbf{Design of low-loss second-order waveguide Bragg grating.} Schematics of a shallow-etched grating (\textbf{a}) and a deep-etched grating (\textbf{b}), both not to scale. \textbf{c}, Simulated radiation loss versus grating coupling coefficient $\kappa$. The curve parameter is the gap between the waveguide and the grating.
    }
    \label{fig:2nd_order_grating_loss}
\end{figure}

Our experimental implementation of the RE-DBR external cavity employs a second-order Bragg grating. Compared to a first-order grating, this design features larger critical dimensions and is therefore more fabrication-friendly. However, higher-order gratings inherently introduce excess radiation loss that can degrade device performance~\cite{streifer1977coupled}. To suppress this radiation loss without increasing the lithographic resolution requirements, we adopt a shallow-etched grating design for our RE-DBR device, as illustrated in Fig.~\ref{fig:2nd_order_grating_loss}a.

This grating is fabricated alongside a 2.8-$\mu$m-wide, 360-nm-thick TFLN ridge waveguide. A shallow etch step leaves a periodic array of 300-nm-diameter, 40-nm-deep holes in the 180-nm-thick residual lithium niobate slab. Our approach contrasts with conventional Bragg gratings based on periodically arranged posts, which are typically formed together with the waveguide in a single etch step~\cite{siddharth2025ultrafast,xiang2019ultra} (see Fig.~\ref{fig:2nd_order_grating_loss}b).

We analyze the radiation characteristics of these higher-order gratings using coupled-wave theory~\cite{zhong2004effect,zhong2004two,millett2008simulation,millett2009modeling,millett2010impact}. A complete theoretical treatment and experimental verification of the grating radiation are presented in a separate publication~\cite{li2026low}. Here, we briefly describe the simulation settings and summarize the key results. We numerically compute the radiation loss and grating coupling coefficient for the shallow-etched and deep-etched designs. To ensure a fair comparison, the post diameter in the deep-etched design is set to 300 nm---the same as the hole diameter in the shallow-etched design. We then sweep the gap between the grating structures and the waveguide and monitor how the radiation loss and coupling coefficient evolve.

Figure~\ref{fig:2nd_order_grating_loss}c compares the simulated radiation loss as a function of the grating coupling coefficient for the two designs. Notably, at the same coupling coefficient $\kappa$, the shallow-etched grating exhibits an 8-fold reduction in radiation loss relative to the deep-etched grating. With this shallow-etched design, our device achieves a coupling coefficient of 0.1 cm\textsuperscript{-1} and a radiation loss of only 2.3 dB\,m\textsuperscript{-1}, enabling strong, narrowband synthetic feedback and, consequently, narrow-linewidth lasing.

\section{Analysis of thermal fluctuations in lithium niobate resonators} \label{SI_tccr_noise}

In silicon nitride resonators, thermo-refractive noise (TRN) is typically the dominant source of thermally driven frequency fluctuations. In lithium niobate resonators, however, frequency stability is primarily limited by thermal-charge-carrier-refractive (TCCR) noise. TCCR noise arises from local thermal fluctuations in the charge-carrier density, which generate local electric-field fluctuations. These electric-field fluctuations are subsequently converted into resonance-frequency noise through the electro-optic effect.

We consider a X-cut lithium niobate racetrack resonator, where the optical-mode polarization is parallel to the material's optical axis along the straight sections. For a resonance at frequency $f_{0}$, the single-sideband power spectral density of the TCCR frequency noise can be approximated as~\cite{zhang2025fundamental}

\begin{equation} \label{eq:tccr_spectrum}
S_{\nu,\mathrm{TCCR}}(f) = \frac{n^{4}r^{2}_{33}k_{B}T}{2\sigma(f)V}\frac{1}{1+\left(2\pi\frac{\epsilon_{0}\epsilon_{r}}{\sigma(f)}f\right)^{2}}f_{0}^{2},
\end{equation}

where $f$ denotes the offset frequency. The parameters in Eq.~\ref{eq:tccr_spectrum}, along with the values used in our numerical simulations, are summarized in Table~\ref{tab:tccr_calc_param}~\cite{mansingh1985ac,ghione2009semiconductor}. Using this parameter set, we obtain a single-sideband TCCR frequency-noise spectral density of 2.2 Hz\textsuperscript{2}\,Hz\textsuperscript{-1} at a 10-MHz offset frequency. This corresponds to a short-term linewidth of 13.8 Hz, as reported in the main text.

\begin{table}[h]
    \centering
    \renewcommand{\arraystretch}{1.25}
    \setlength{\tabcolsep}{6pt}
    \small
    
    \begin{tabular}{|c|c|c|c|}
    \hline
    \multicolumn{1}{|c|}{\textbf{Symbol}} &
    \textbf{Value} &
    \textbf{Unit} &
    \textbf{Definition} \\
    \hline

    $n$ & 2.1376 & 1 & Refractive index of lithium niobate along the extraordinary axis \\
    $r_{33}$ & $30.5\times10^{-12}$ & m\,V\textsuperscript{-1} & Pockels coefficient \\
    $k_{B}$ & $1.38\times10^{-23}$ & J\,K\textsuperscript{-1} & Boltzmann constant \\
    $T$ & 298 & K & Ambient temperature \\
    $\sigma(f)$ & $5\times10^{-9}\times\left(\frac{f}{\mathrm{1~kHz}}\right)^{0.8}$ & S\,m\textsuperscript{-1} & Electrical conductivity of lithium niobate at offset frequency $f$ \\
    $V$ & $1.84\times10^{-14}$ & m\textsuperscript{3} & Volume of the resonator \\
    $\epsilon_{0}$ & $8.854\times10^{-12}$ & F\,m\textsuperscript{-1} & Vacuum permittivity \\
    $\epsilon_{r}$ & 29 & 1 & Dielectric constant of lithium niobate \\
    $f_{0}$ & 193.4 & THz & Resonance frequency \\
    
    \hline
    \end{tabular}
    
    \caption{\textbf{Parameters used in the simulations of TCCR frequency noise spectrum.}}
    \label{tab:tccr_calc_param}

\end{table}


\section{Phase-noise-limited ranging precision of FMCW LiDAR} \label{SI_ranging_precision}

The ranging precision of a FMCW LiDAR is constrained by the phase noise in the laser output. Such phase noise includes contributions of the spontaneous emission, thermal noise, and the RF signal source that drives the frequency modulation. In this section, we establish the laser phase noise spectrum $S_{\phi}(f)$, or equivalently the frequency noise spectrum $S_{\nu}(f) = \nu^{2}S_{\phi}(f)$, and the corresponding FMCW LiDAR ranging error. In general, this phase-noise-limited ranging precision of the FMCW LiDAR system is derived by treating the measured ranging as a finite-time average of the heterodyne beat frequency.

Consider a laser under linear frequency modulation, its instantaneous frequency is given by

\begin{equation}  \label{eq:instant_laser_freq}
    \nu(t) = \nu_{0} + \gamma t + \delta\nu(t),
\end{equation}

where $\gamma$ is the nominal chirp rate in Hz\,s\textsuperscript{-1}, and $\delta\nu$ is the zero-mean laser frequency fluctuation. For a target at distance $L$, the round-trip delay is $\tau_{0} = \frac{2L}{c}$. The instantaneous FMCW beat-note frequency is then

\begin{equation}  \label{eq:instant_beat_freq}
    \nu_{b}(t) = \nu(t) - \nu(t-\tau_0).
\end{equation}

For an ideal linear chirp, this gives the deterministic beat frequency $\gamma\tau_{0}$. The residual fluctuating component is

\begin{equation}  \label{eq:fluctuate_beat_freq}
    \delta\nu_{b}(t) = \delta\nu(t) - \delta\nu(t-\tau_0).
\end{equation}

In our LiDAR demonstration, the beat frequency is estimated by averaging over an integration time $w$, the corresponding range estimate is

\begin{equation}  \label{eq:range_estimate}
    \hat{L}(t) = \frac{c}{2\gamma w}\int_{t-w}^{t}\nu_{b}(s)ds.
\end{equation}

Therefore, the phase-noise-induced range error is

\begin{equation}  \label{eq:range_error}
    \delta L(t) = \frac{c}{2\gamma w}\int_{t-w}^{t}[\delta\nu(s) - \delta\nu(s-\tau_{0})]ds.
\end{equation}

This equation shows that laser frequency fluctuation is converted to range noise through two filtering operations. First, the finite averaging time $w$ introduces a low-pass response. Second, the delayed self-difference $\delta\nu(t) - \delta\nu(t-\tau_{0})$ introduces a high-pass response, because slow frequency fluctuations are nearly common-mode over the round-trip delay and are therefore suppressed.

Using the Fourier transform convention

\begin{equation}  \label{eq:fourier_transform}
    X(f) = \int_{-\infty}^{\infty}x(t)e^{-i2\pi ft}dt,
\end{equation}

the transfer function from laser fluctuation $\delta\nu(t)$ to range error $\delta L(t)$ is

\begin{equation}  \label{eq:range_error_transfer_func}
    H_{L}(f) = \frac{c}{2\gamma}\mathrm{sinc}(\pi fw)(1-e^{-i2\pi f\tau_{0}}),
\end{equation}

where $\mathrm{sinc}(x) = \sin(x)/x$. Let $S_{\nu}(f)$ denote the single-sideband power spectral density of the laser frequency noise, with units of Hz\textsuperscript{2}\,Hz\textsuperscript{-1}. The corresponding single-sideband power spectral density of the range error is

\begin{equation}  \label{eq:range_error_spectrum}
    S_{L}(f) = |H_{L}|^{2}S_{\nu}(f) = \left(\frac{c}{2\gamma}\right)^{2}\mathrm{sinc}^{2}(\pi fw)[2 - 2\cos(2\pi f\tau_{0})]S_{\nu}(f).
\end{equation}

Therefore, the phase-noise-induced rms range error is

\begin{equation}  \label{eq:range_error_from_phase_noise}
    \boxed{\sigma_{L} = \left[\int_{-\infty}^{\infty}S_{L}(f)df\right]^{1/2} = \frac{c}{2\gamma}\left[\int_{-\infty}^{\infty}\mathrm{sinc}^{2}(\pi fw)[2 - 2\cos(2\pi f\tau_{0})]S_{\nu}(f)df\right]^{1/2},}
\end{equation}

with $\tau_{0} = 2L/c$ the photon roundtrip time and $w$ the integration time. This expression is the desired relationship between FMCW LiDAR ranging error and laser phase noise. The phase-noise-limited RMS ranging precision shown in the main text is obtained by inserting the laser frequency noise spectrum using Eq.~\ref{eq:range_error_from_phase_noise}. Notably, this laser frequency noise spectrum is measured when the laser undergoes slow frequency modulation, rather than under the static conditions, to include the phase noise from the RF source.

\section{Theory of fiber-optic acoustic sensing} \label{SI_fiber_sensing}

In this section, we show that in optical frequency-domain reflectometry (OFDR), dynamic fiber strain manifests as a phase modulation of the electrical beat note. By tracking the phase of the corresponding Fourier coefficient from one laser ramp to the next, the dynamic strain can be extracted.

The OFDR experimental setup is shown in Fig.~5a of the main text. A piezoelectric transducer (PZT) induces strain in an active section of the sensing fiber, thereby generating an acoustic perturbation. If the strain is uniform over the active sensing length $L_s$, denoted by $\epsilon(t)$, the corresponding physical elongation is
\begin{equation}
    \Delta L(t)=L_s\epsilon(t).
\end{equation}
The associated round-trip optical phase change is
\begin{equation}
    \Delta\phi_{\epsilon}(t)=\frac{4\pi}{\lambda}n_{\rm eff}(1-p_e)L_s\epsilon(t),
    \label{eq:strain_induced_phase_change_raw}
\end{equation}
where $p_e$ is the effective photoelastic coefficient, $n_{\rm eff}$ is the effective refractive index, and $\lambda$ is the optical wavelength.

We next describe how this acoustic phase modulation is encoded in the OFDR beat note. During each linear frequency ramp of the FMCW laser, the local oscillator field can be written as
\begin{equation}
    E_{\rm LO}(t)=E_{\rm LO}\exp\left[j2\pi\left(\nu_0 t+\frac{1}{2}\gamma t^2\right)\right],
    \label{eq:local_oscillator_field}
\end{equation}
where $\nu_0$ is the initial optical frequency of the ramp and $\gamma$ is the frequency tuning rate. The reflected field from a reflector located at distance $L$ is delayed by the round-trip time
\begin{equation}
    \tau=\frac{2n_gL}{c},
\end{equation}
where $n_g$ is the group refractive index. Including the strain-induced phase change, the reflected field is
\begin{equation}
    E_r(t)=\rho E_0\exp\left[j2\pi\left(\nu_0(t-\tau)+\frac{1}{2}\gamma(t-\tau)^2\right)+j\Delta\phi_{\epsilon}(t)\right],
    \label{eq:reflected_field}
\end{equation}
where $\rho$ is the reflector amplitude coefficient and $E_0$ is the input field amplitude.

The detected electrical beat note is proportional to the interference term $E_{\rm LO}^{*}(t)E_r(t)$. Therefore,
\begin{equation}
    s(t)\propto E_{\rm LO}^{*}(t)E_r(t)\propto \exp\left[-j2\pi f_b t-j\phi_0+j\Delta\phi_{\epsilon}(t)\right],
    \label{eq:ofdr_beatnote}
\end{equation}
where
\begin{equation}
    f_b=\gamma\tau
\end{equation}
is the OFDR beat frequency and
\begin{equation}
    \phi_0=2\pi\left(\nu_0\tau-\frac{1}{2}\gamma\tau^2\right)
\end{equation}
is a static phase offset. Equation~\eqref{eq:ofdr_beatnote} shows that the dynamic strain does not primarily change the beat frequency; instead, it modulates the phase of the beat note.

Because the FMCW laser is driven by triangular-wave frequency modulation, the measured beat note is naturally divided into individual linear frequency ramps. For the $m$-th ramp, we introduce a fast time $t$ within the ramp and a slow-time index $t_m$ labeling the ramp. If the acoustic strain varies slowly compared with the duration of one ramp, the strain-induced phase can be treated as approximately constant during that ramp:
\begin{equation}
    \Delta\phi_{\epsilon}(t)\approx\Delta\phi_{\epsilon}(t_m).
\end{equation}
The beat note within the $m$-th ramp is then
\begin{equation}
    s_m(t)\propto\exp\left[-j2\pi f_b t-j\phi_0+j\Delta\phi_{\epsilon}(t_m)\right].
    \label{eq:ofdr_beatnote_mth_ramp}
\end{equation}

Applying a FFT to the beat note segment from the $m$-th ramp produces a Fourier component near $-f_b$, corresponding to the reflector at distance $L$. The complex Fourier coefficient at this component can be written as
\begin{equation}
    S_m(-f_b) = A_m\exp\left[-j\phi_0+j\Delta\phi_{\epsilon}(t_m)\right],
    \label{eq:ofdr_fft_coefficient}
\end{equation}
where $A_m$ accounts for the reflection amplitude, windowing function, and finite FFT-bin response. If these contributions are constant, the phase of the Fourier coefficient directly tracks the strain-induced phase:
\begin{equation}
    \mathrm{Arg}\left[S_m(-f_b)\right]=\Delta\phi_{\epsilon}(t_m)-\phi_0',
    \label{eq:ofdr_fft_phase}
\end{equation}

where $\phi_{0}' = \phi_{0} - \mathrm{Arg}(A_{m})$ is a constant phase term. Combining Eq.~\eqref{eq:ofdr_fft_phase} with Eq.~\eqref{eq:strain_induced_phase_change_raw}, the dynamic strain sampled once per laser ramp is obtained as
\begin{equation}
    \boxed{\epsilon(t_m)=\frac{\lambda}{4\pi n_{\rm eff}(1-p_e)L_s}\left[\mathrm{Arg}\left[S_m(-f_b)\right]+\phi_0\right].}
    \label{eq:ofdr_fiber_strain}
\end{equation}

Thus, the acoustic perturbation is recovered by monitoring the ramp-to-ramp phase variation of the selected OFDR beat note Fourier coefficient. In practice, the constant conversion factor and static phase offset can be determined through calibration measurements. Phase unwrapping may also be applied to obtain a continuous strain waveform when the dynamic phase excursion exceeds the principal phase range.

\section{Characterization of waveguide Bragg grating and E-DBR laser} \label{SI_edbr_laser}

In the main text, we compared the frequency noise characteristics of our RE-DBR laser with those of an E-DBR laser that uses an 8-mm-long waveguide Bragg grating as an external cavity. To support this comparison, we provide here additional characterization results for both the Bragg grating and the E-DBR laser.

\begin{figure}[t!]
    \centering
    \includegraphics{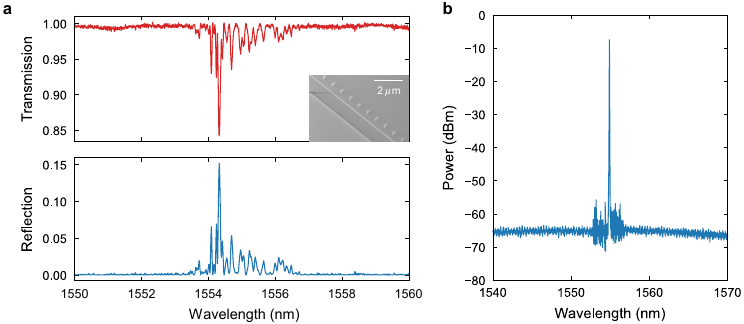}
    \captionsetup{singlelinecheck=off, justification = RaggedRight}
    \caption{
        \textbf{Characterization of waveguide Bragg grating and E-DBR laser.} \textbf{a}, Transmission and reflection spectra of the Bragg grating. The inset shows a SEM image of the device. \textbf{b}, Optical spectrum of an E-DBR laser consisting of a RSOA and the grating as an external cavity, measured at an injection current of 80 mA, corresponding to the laser frequency noise spectrum shown in the main text.
    }
    \label{fig:edbr_laser}
\end{figure}

The waveguide Bragg grating comprises an array of shallow-etched holes, each 300 nm in diameter and 40 nm deep. These holes are arranged with a period of 876 nm, offset 550 nm from a 1-$\mu$m-wide ridge waveguide (Fig.~\ref{fig:edbr_laser}, inset), forming a second-order grating with a Bragg wavelength near 1550 nm. The transmission and reflection spectra of the device are shown in Fig.~\ref{fig:edbr_laser}a, with peak reflectivity occurring at 1554 nm.

The E-DBR laser is formed by butt-coupling this Bragg grating to a RSOA, where the grating acts as the external cavity. Figure~\ref{fig:edbr_laser}b displays the optical spectrum of this laser at an injection current of 80 mA; the laser achieves a side-mode suppression ratio (SMSR) of approximately 50 dB. The frequency noise power spectral density of the E-DBR laser presented in the main text corresponds to measurements taken under these same operating conditions.

\section{Characterization of small-signal modulation response of RE-DBR} \label{SI_s21_measurement}

\begin{figure}
    \centering
    \includegraphics{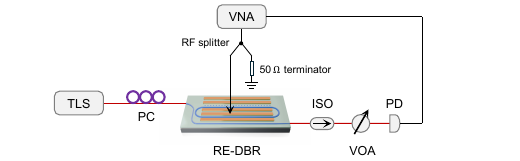}
    \captionsetup{singlelinecheck=off, justification = RaggedRight}
    \caption{
        \textbf{Experimental setup for electro-optic modulation response $S_{21}$ measurement.}
        This setup comprises a tunable laser source (TLS), a vector network analyzer (VNA), a polarization controller (PC), a fiber-optic isolator (ISO), a RF power splitter, a variable optical attenuator (VOA), and a photodetector (PD). The TLS wavelength is set to the linear region of a RE-DBR resonance, such that the resonant frequency shift induced by electro-optic modulation produces a directly proportional change in the transmitted optical power.
    }
    \label{fig:S21_measurement_setup}
\end{figure}

The small-signal modulation response was obtained by measuring the voltage-to-transmission \(S_{21}\) of the RE-DBR external cavity, following the standard characterization procedure for microring intensity modulators. A schematic of the measurement setup is shown in Fig.~\ref{fig:S21_measurement_setup}; it includes a tunable laser source (TLS, TOPTICA CTL 1550), a vector network analyzer (VNA, Keysight N4372E-476 PNA-X), a RF power splitter, a 50-\(\Omega\) terminator, and a high-bandwidth photodetector (PD, Thorlabs PDB480C-AC). The TLS was tuned to the linear region of a RE-DBR resonance. The two output ports of the power splitter were connected to a 50-\(\Omega\) terminator and a high-impedance load (the modulator electrode), respectively, to achieve impedance matching and minimize RF return loss. To simplify the measurement, the RF signal from the power splitter was applied to only one signal electrode of the RE-DBR, leaving the other electrode idle. This measurement yielded the intensity modulation response shown in Fig.~3c of the main text, from which the 3-dB intensity modulation bandwidth was extracted.

\medskip

\clearpage

\bibliography{main.bib}